\documentclass[journal]{IEEEtran}
\usepackage{graphicx}
\usepackage{epsfig,cite,amsmath,amssymb,latexsym,subfigure}
\usepackage{floatflt}
\usepackage{theorem}
\usepackage{url}
\usepackage{times}
\usepackage[usenames]{color}
\newcommand{\E}[1]{E\left[#1\right]}
\newcommand{\set}[1]{\left[#1\right]}

\title{\Large \bf Doped Fountain Coding for Minimum Delay Data Collection in Circular Networks}

\author{\normalsize Silvija Kokalj-Filipovi\'c, Predrag Spasojevi\'c\thanks{S. Kokalj-Filipovi\'c and P. Spasojevi\'{c} are with Wireless Information
Network Laboratory (WINLAB), Department of Electrical and Computer
Engineering, Rutgers University,
North Brunswick, NJ 08902, USA (e-mail:
\{skokalj,spasojev\}@winlab.rutgers.edu).}%
, and Emina Soljanin\thanks{E. Soljanin is with Lucent-Alcatel, Bell
Labs, Murray Hill, NJ, USA (email: emina@research.bell-labs.com).}}

\newcommand{\prob}[1]{\mbox{Pr}\left\{#1\right\}}

\newcommand{\eqnref}[1]{(\ref{eqn:#1})}

\newcommand{\eqnlabel}[1]{\label{eqn:#1}}

\newcommand     {\paren}[1]{\left(#1\right)}

\newcommand{\curlb}[1]{\left\{#1\right\}}

\newcommand{\eX}[1]{\e^{#1}}

\newcommand{\e}{e}

\begin{document}
\renewcommand\topfraction{.65} 
\renewcommand\textfraction{.35}
\date{}
\maketitle
\footnote{This paper is in part based on \cite{CissFountain} and \cite{ISSSTAFountain}}
\begin{abstract}
This paper studies decentralized, Fountain and network-coding based
strategies for facilitating data collection in circular wireless
sensor networks, which rely on the stochastic diversity of data
storage. The goal is to allow for a reduced delay
 collection by a data collector who accesses the network at a
 random position and random time. Data dissemination is performed
by a set of relays which form a circular route to exchange source
packets. The storage nodes within the transmission range of the
route's relays linearly combine and store overheard relay
transmissions using random decentralized strategies. An intelligent
data collector first collects a minimum set of coded packets from a
subset of storage nodes in its proximity, which might be sufficient
for recovering the original packets and, by using a message-passing
decoder, attempts recovering all original source packets from this
set. Whenever the decoder stalls, the source packet which restarts
decoding is polled/doped from its original source node. The
random-walk-based analysis of the decoding/doping process furnishes
the collection delay analysis with a prediction on the number of
required doped packets. The number of doped packets can be
surprisingly small when employed with an Ideal Soliton code degree
distribution and, hence, the doping strategy may have the least
collection delay when the density of source nodes is sufficiently
large. Furthermore, we demonstrate that network coding makes
dissemination more efficient at the expense of a larger collection
delay.
Not surprisingly,  a circular network allows for a 
significantly more (analytically and otherwise) tractable strategies
relative to a network whose model is a random geometric graph.
\end{abstract}
{\bf Keywords:}
decentralized Fountain codes, wireless networks, network coding, distributed storage, data collection 
  \section{Introduction} \label{sec:intro}
{\em Wireless sensor networks (WSN)}  monitor and collect  sensor
data distributed over large physical areas. Sensor nodes are simple,
battery-run devices with limited data processing, storage, and
transmission capabilities.
  For energy efficiency reasons, the main
data propagation model is hop-by-hop, where nodes relay other nodes'
data.  An important collection scenario is when 
a data sink (a {\em collector}) appears at a random position, at random time, and aims
to collect  all the $k$ source data packets.
The network's goal is to ensure  that the  data packets be
efficiently disseminated and stored in a manner which allows for a
low collection delay upon collector's arrival. This is achieved by
storing data
 in a compact collection area
{\em at the fingertips} of the data collector, i.e., at a set
of connected (through multiple hops) wireless nodes in its
proximity.  There is a fundamental tradeoff between network
storage capacity and the collection delay. If each node across the network can store all the packets, the data can be collected in
a single hop. On the other extreme, if each node has own data
destined for the collector and no capacity to store other packets, the collector has to reach out to all the network nodes
to collect the data, which would incur an extreme delay. The
canonical model considered here is when the number of source nodes $k$ is smaller than the number of network nodes
and where each network node can both relay and store one packet.

Storing data at the fingertips of the randomly positioned collector
implies redundant data storage across the network, whether it means
simply storing source packet replicas, or random linear combinations
thereof, and resulting respectively in a repetition code, or other
linear code, implemented across a network. More efficient storage
codes than the simplest repetition code require that the collector
not only collects the linear combinations
 but also is capable of decoding/recovering the original
packets.
The two general classes of  packet combining (coding)
techniques in~\cite{DimakisRamchandran,DimakisFountain,CissFountain,SalahEmina}, 
and \cite{LiLiLi} are: Fountain-type erasure  codes 
\cite{LiLiLi,DimakisFountain,CissFountain,SalahEmina}, 
and decentralized erasure codes \cite{DimakisRamchandran} - a
variant of random linear network
codes~\cite{RandCodBenefit,RandNetCoding}. The advantage of Fountain
type coding is in the linear complexity of decoding which, here,
corresponds to linear original packet recovery time. 
The key difficulty of the Fountain type
  storage approach is in devising efficient
techniques to disseminate data from multiple sources to network
storage nodes in a manner which ensures that the required statistics
of  created linear combinations is accomplished. Achieving this goal
is  particularly difficult when employed with the classic random
geographic graph network models~\cite{LiLiLi,SalahEmina}.

\begin{figure}[!t] 
\begin{center}
\includegraphics[width=3.7in]{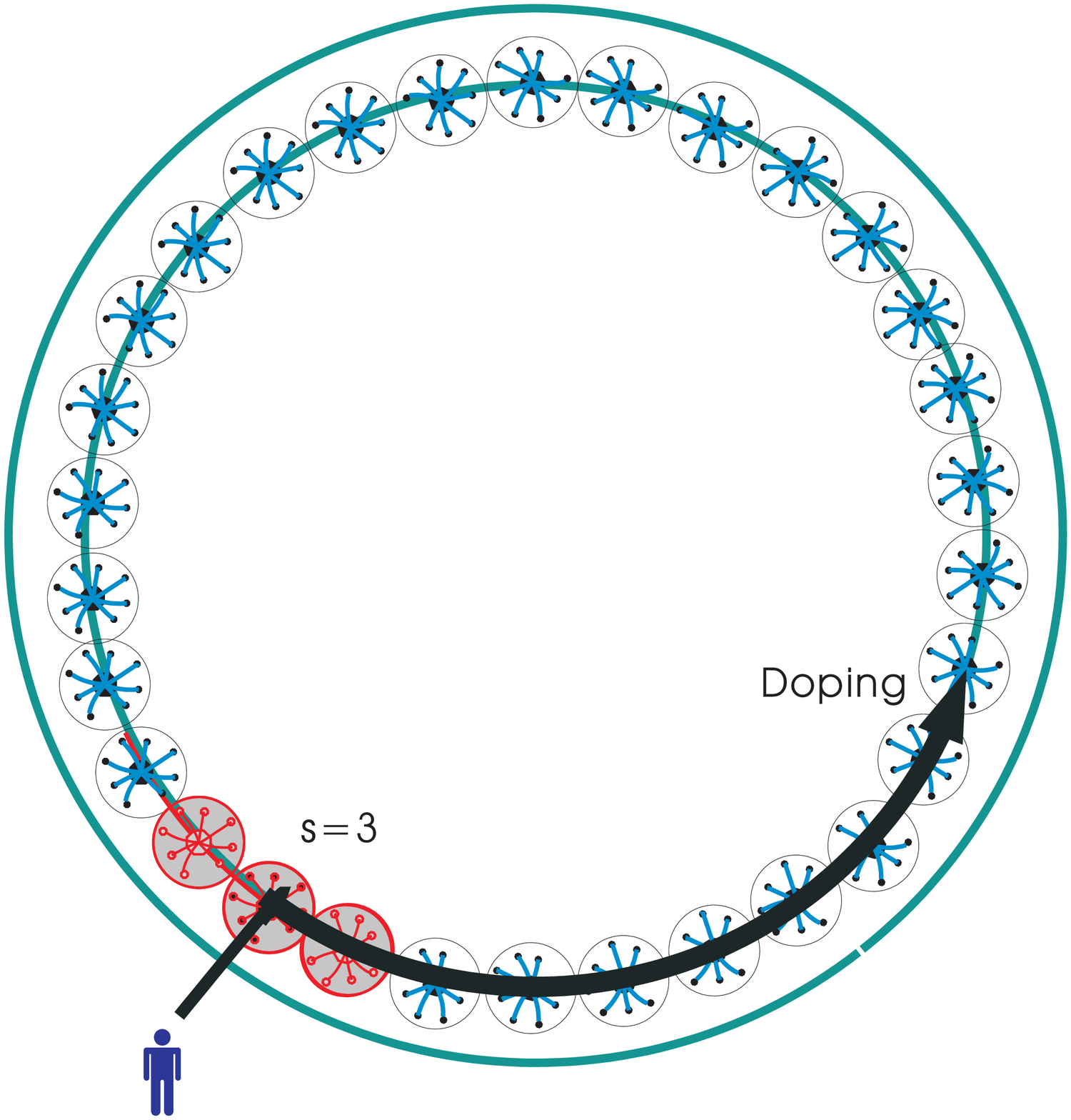}
\caption{Collection of coded symbols: pull phase brings the three squads of coded packets to the decoder, and then, whenever the decoder gets stalled, an original symbol is pulled off the network for doping. We here deliberately omit to show that the squad nodes are overhearing (belonging to) two adjacent relays in order to highlight the two-phase collection, as opposed to the storage protocol.}
\label{fig:pushpullcoll}
\end{center}
\end{figure}

In this paper we analyze decentralized Fountain-type network coding
strategies for facilitating a reduced delay data collection and
network coding schemes for efficient data dissemination for 
a planar donut-shaped sensor network (see Fig.~\ref{fig:pushpullcoll}) whose nodes lie
between two concentric circles. 
The network backbone is a circular route of relay nodes which disseminate data. All network
nodes within its transmission range overhear relay's transmissions and serve as potential storage nodes. The storage
nodes within a relay's transmission range form a squad. The squad
size  determines the relay's one-hop storage capacity. Squad's
storage capacity together with the source node density and the
coding/collection strategy determine the data collection delay
measured in terms of the number of communication hops required for
the collector to collect and recover all $k$
source data packets. 
In the proposed polling (packet doping) scheme, an intelligent
data collector {\em (IDC)} first collects a minimum set of coded packets from a
subset of storage squads in its proximity (as in Fig.~\ref{fig:pushpullcoll}), which might be sufficient
for recovering the original packets and, by using a message-passing
decoder, attempts recovering all original source packets from this
set. Whenever the decoder stalls, the source packet which
restarts decoding is polled/doped from its original source node (at
an increased delay since this packet is likely not to be close to the
collector).
 The random-walk-based analysis of the
decoding/doping  process represents the key contribution of this
paper. It furnishes the collection delay analysis with a prediction
on the number of required doped packets.
The number of required packet dopings is surprisingly small and,
hence, to reduce the number of collection hops required to recover the source data, one should employ the  {\em doping}
collection scheme. The delay gain due to doping is more significant
when the relay squad storage capacity is smaller. Furthermore,
employing network coding makes dissemination more efficient at the
expense of a larger collection delay.

Not surprisingly,  a circular network allows for a 
significantly more (analytically) tractable strategy  relative to
a network whose model is a random geometric
graph ({\em RGG})~\cite{LiLiLi,SalahEmina}. Besides, the RGG modeling implies that a packet is forwarded to one of the neighbors in the network
graph, while the fact that all the neighbors are overhearing the same transmission
is not considered. In contrast, the proposed approach is aiming to
incorporate the {\em wireless multicast advantage}~\cite{BIPwlessADV} into the dissemination/storage model. In our earlier work \cite{InfrastrProp}, we show how a randomly deployed network can self-organize into concentric donut-shaped networks. Note that the proposed topology is an especially good model for sensor networks deployed to monitor physical phenomena with linear spatial blueprint, such as road networks, vehicular networks, or border and pipeline-security sensor nets \cite{eriksson2008pothole,MoteView}.

  \section{System Model and Problem Formulation} \label{Sect:Motiv}
\begin{figure}[!t] 
\begin{center}
\includegraphics[width=3.9in]{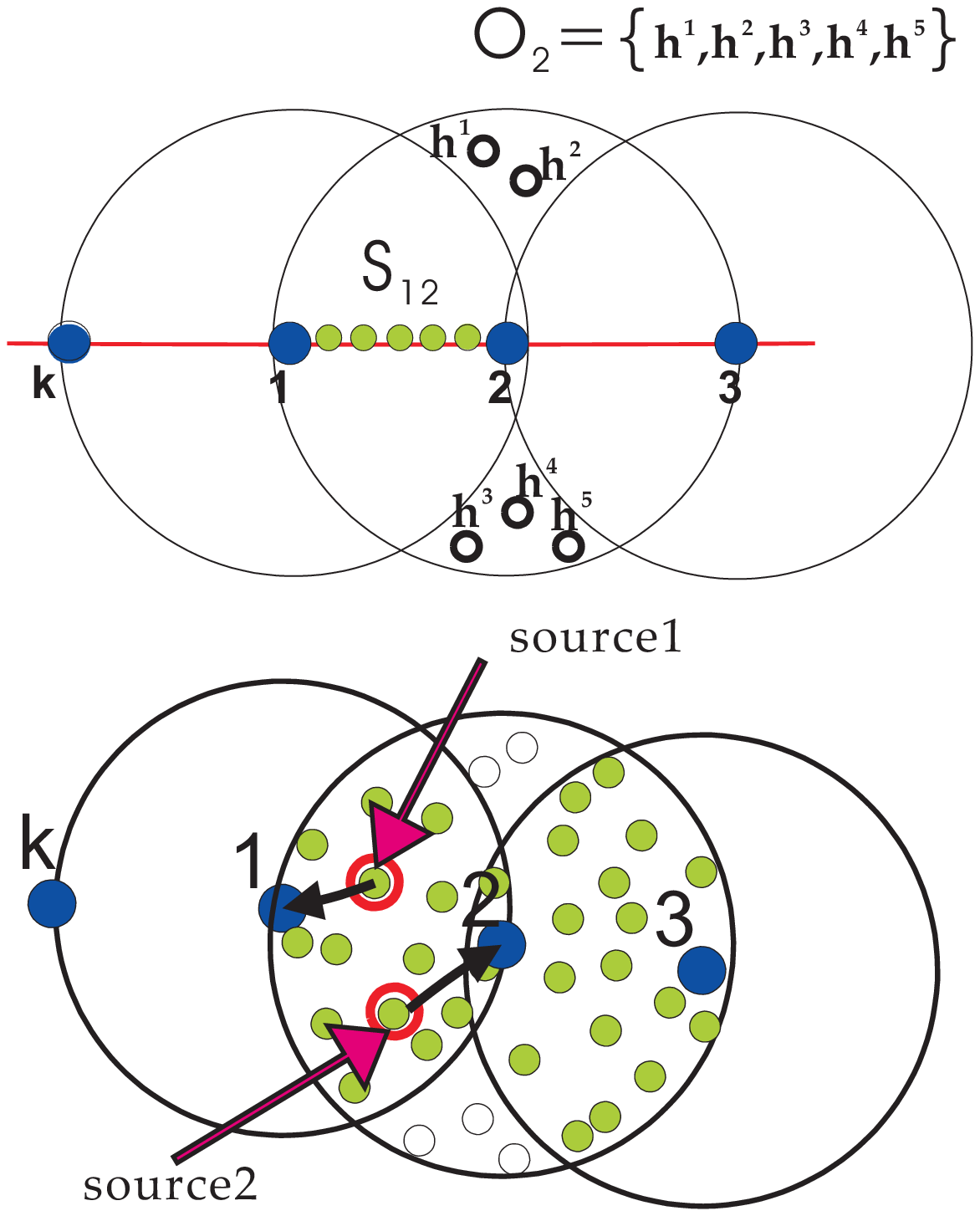}
\caption{Close-Up of a Circular Squad Network of $k$ relays. We assume there is one source per relay on average - the source entrusts its data packet to the closest relay, hence making it a virtual source. Each relay is overheard by nodes in its transmission range, referred to as squad nodes.}
\label{fig:closupa}
\end{center}
\end{figure}

We consider an inaccessible static wireless sensor network (e.g., a
disaster recovery network) with network nodes that are capable of
sensing, relaying, and storing data. Nodes are randomly scattered in
a plane according to a Poisson point process of some intensity
$\mu$. The nodes have constrained memory resources. Without loss
of generality, we assume that most nodes have a unit-size buffer.
Each node that senses an event creates a unit-size description data packet. We refer to such a node as a data source. We assume
that events are distributed as a Poisson point process of intensity
$\mu_s<\mu$. We define the transmission range as the maximum
distance $r$ from the transmitter at which nodes can reliably
receive a packet. Assuming radially symmetric attenuation (isotropic
propagation), 
 the transmitted packet is reliably
received in a disk of area $r^2\pi,$  illustrated
in~Figure~\ref{fig:closupa}. The expected number of network nodes in
the disk is $\mu r^2\pi$ and the expected number of source nodes is $\mu_s r^2\pi.$

Within the sensor network, we consider a circular route, composed of
$k$ nodes referred to as relays. 
The distance between adjacent relays is equal to the transmission
range $r$. Without a loss of generality and for simplicity, we
assume that $r$ is selected so that only a single data source node
is (expected to be) within the transmission range of a relay. That
is, $\mu_s r^2\pi=1.$ Source node observes an event and sends
its data packet to the relay, making it a virtual source (see
Figure~\ref{fig:closupa}). Thus, $k$ relays form a linear network
(route) with data packet $i$ assigned to relay $i,$
$i\in\set{1,\ldots,k}.$ 
Each sensor node within the range of a relay is associated with the
route via a one-hop connection to a relay. We refer to the set of
nodes within the range of a relay as a {\em squad}, and to a node as
{\em squad-node}. Squad nodes can hear transmissions either from
only relay $i$ or from, also relay $i+1$ and, thus, belong to either
the {\em own set of squad-nodes} $O_i,$ or to the {\em shared set of
squad-nodes,} denoted $S_{i(i+1)},$ where, hereafter, any addition
operation will be assumed to be mod $k$, i.e., $(i+1)$~mod~$k$, as
shown in Figure~\ref{fig:closupa}. By means of associations, the
relays in the circular route together with the squad nodes form a
donut-shaped circular squad network.  The expected number of nodes
in the squad is denoted with $h = \mu r^2\pi,$ while the
expected area of each shared set is $\E{S_{i(i+1)}}=\tilde{h}=0.4h.$ 
We primarily focus on shared
squad nodes. In the rest of the paper, whenever we
refer to squad-nodes, we mean shared nodes, and for simplicity we assume $h=\tilde{h}$. 
The goal is to disseminate data from all sources and store them at
squad nodes so that a collector can recover all $k$
original packets with minimum delay. An IDC collects data via a {\em
collection relay}. The data is collected from $k_T(k_d)=k_s+k_d$ storage nodes of
which most $(k_s>>k_d)$ reside in a set of $s$ {\em adjacent} squads, including the collection relay squad. These $s$ squads form a
{\em supersquad} (See Figure~\ref{fig:pushpullcoll}). The number of packets not collected from the supersquad is denoted $k_d.$

Note that the  density of sources $\mu_s$ is dependent on the
spatial characteristics of the monitored physical process, i.e., the
spatial density of events. A well designed sensor network will
ensure that the spatial density of nodes $\mu$ is designed to
properly cover this process. When $r$ is selected to ensure
$r^2\pi\mu_s=1$ then $h=\mu/\mu_s$ is the coverage
redundancy factor.
Furthermore, for a given received  signal-to-noise ratio, the
one-hop transmission energy $E_1$ and the single hop delay $\tau_1$
are inversely proportional to $\mu_s.$  And, for a given
circular route radius $R$, the (expected) number of relays is $k=
R/r.$ 
Hence, for a given transmission range $r$ (or $\mu_s$), the only degree
of freedom is the coverage redundancy factor $h$ (squad size), i.e,
the network density $\mu.$ By reducing $\mu,$ we decrease
the average number of nodes in a squad $h$. This has implications to
the collection (delay and energy) cost. 
The supersquad consists of $s=\left\lceil k_s/h\right\rceil$ squads,
and the average number of hops a packet makes until it is collected
by the IDC is $(s-1)/4+1.$ Hence, the smaller the $\mu,$
 the larger the average collection delay $\tau_s=k_s ((s-1)/4+1) \tau_1$ and the energy
$E_s = k_s \paren{(s-1)/4+1} E_1$ from the supersquad. Henceforth, we will,
without loss of generality, normalize $\tau_1=1$ and $E_1=1.$  The
key collection performance measure will be the average number of
collection hops per source packet $c,$ where  $c=k_s \set{1+(s-1)/4}/k$
when all collected packets are from the supersquad, i.e., $k_T(0)=k_s.$

We will comparatively consider two classes of storage/encoding strategies: in the
first, the IDC collects the original
packets, while in the second one the collected packets are linear
combinations of the original packets and, hence, the IDC needs to decode them to recover source packets. When combining is employed,
constrained by the collection delay, we consider only storage
strategies which allow for decoding methods of linear complexity,
i.e., the use of belief propagation (BP) iterative decoders. Taking as
a reference the case where original packets are encoded into coded
packets whose linear combination degrees follow the Robust Soliton distribution, as in~\cite{LiLiLi}, based on the asymptotic analysis
of LT codes~\cite{luby}, we expect that $k_T(0)=k_s=
k+\sqrt k \log^2(k/\epsilon)$ collected code symbols are required to
decode $(1-\epsilon)k$ original symbols, where $\epsilon$ is a
sufficiently small constant. 
Here the number of collected packets is significantly larger than
$k$ for small to medium number of sources $k$. Hence, collection of this many
packets can be expensive, in particular when the event coverage redundancy
factor $h$ is small. Collecting a smaller number of
packets {\em upfront} would result in a stalled decoding process.
Here, we take advantage of the availability of additional replicas
of source packets along the circular network, to pull one such
packet off the network in order to continue the
stalled decoding process. 
See Figure~\ref{fig:pushpullcoll}. The pull phase is meant to assist
the decoding process using a technique that we refer to as {\em
doping}. In the following, encoding  describes the mapping on
the source packets employed both while disseminating and while storing. It is a
mapping from the original $k$ packets to the collected $k_T(k_d)=k_s+k_d$
encoded packets.
\section{Data Dissemination} \label{sect:Dissem}
\begin{figure}[!t] 
\begin{center}
\includegraphics[width=3.9in]{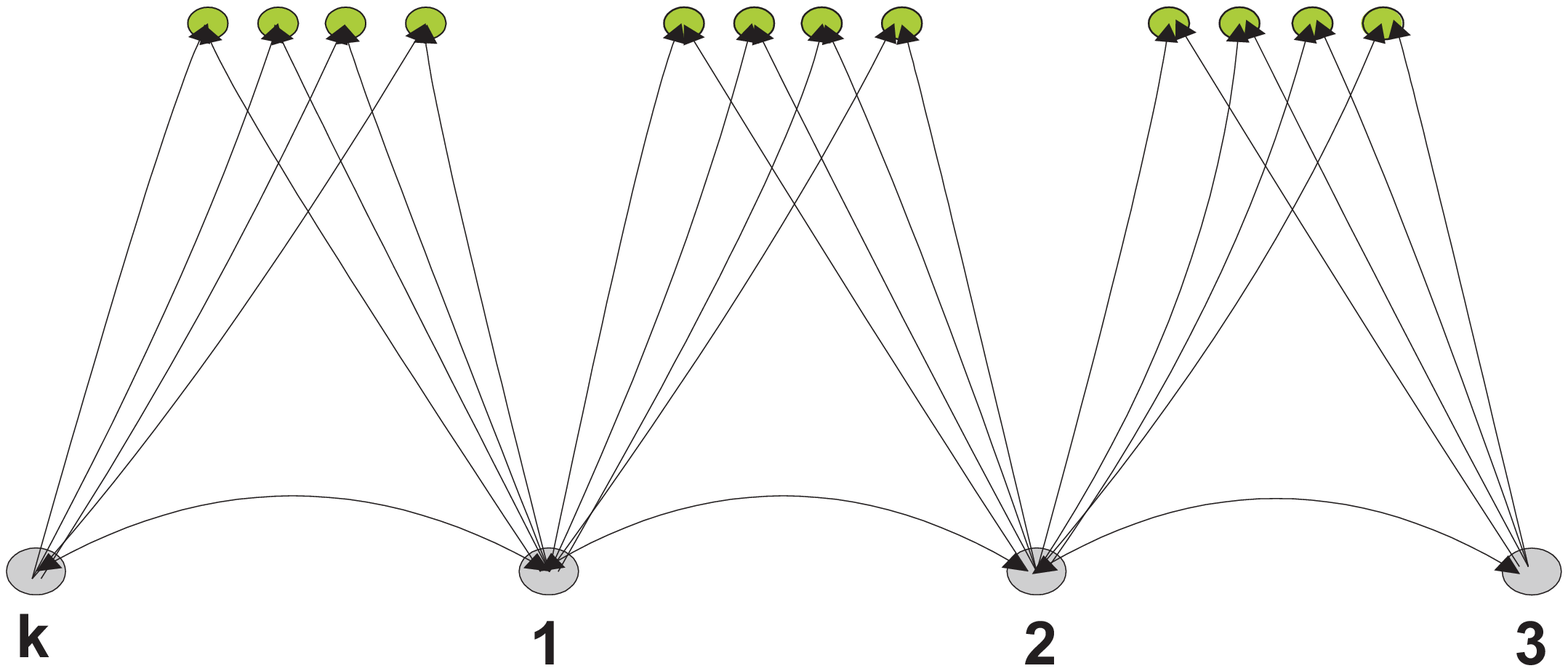}
\caption{Circular Squad Network: the storage graph.}
\label{fig:closupb}
\end{center}
\end{figure}

The nodes within the
transmission range of the route relays together with the relays
themselves form a dissemination network. 
The dissemination connectivity graph
 is a simple circular graph with $k$ nodes. This graph
 models connections between relays, which are bidirectional. The
connectivity graph used in the storage model is  expanded  with
storage nodes, representing shared squad nodes. 
In this graph, every storage node is adjacent to two neighboring
relay nodes. 
Also, edges between storage and relay nodes are directed, as
illustrated in Figure~\ref{fig:closupb}.
 Every edge in the dissemination graph is of unit capacity. 
 A single transmission reaches two neighboring  relays.

\begin{figure}[!t] 
\begin{center}
\hspace{-1cm}
\includegraphics[width=3.9in]{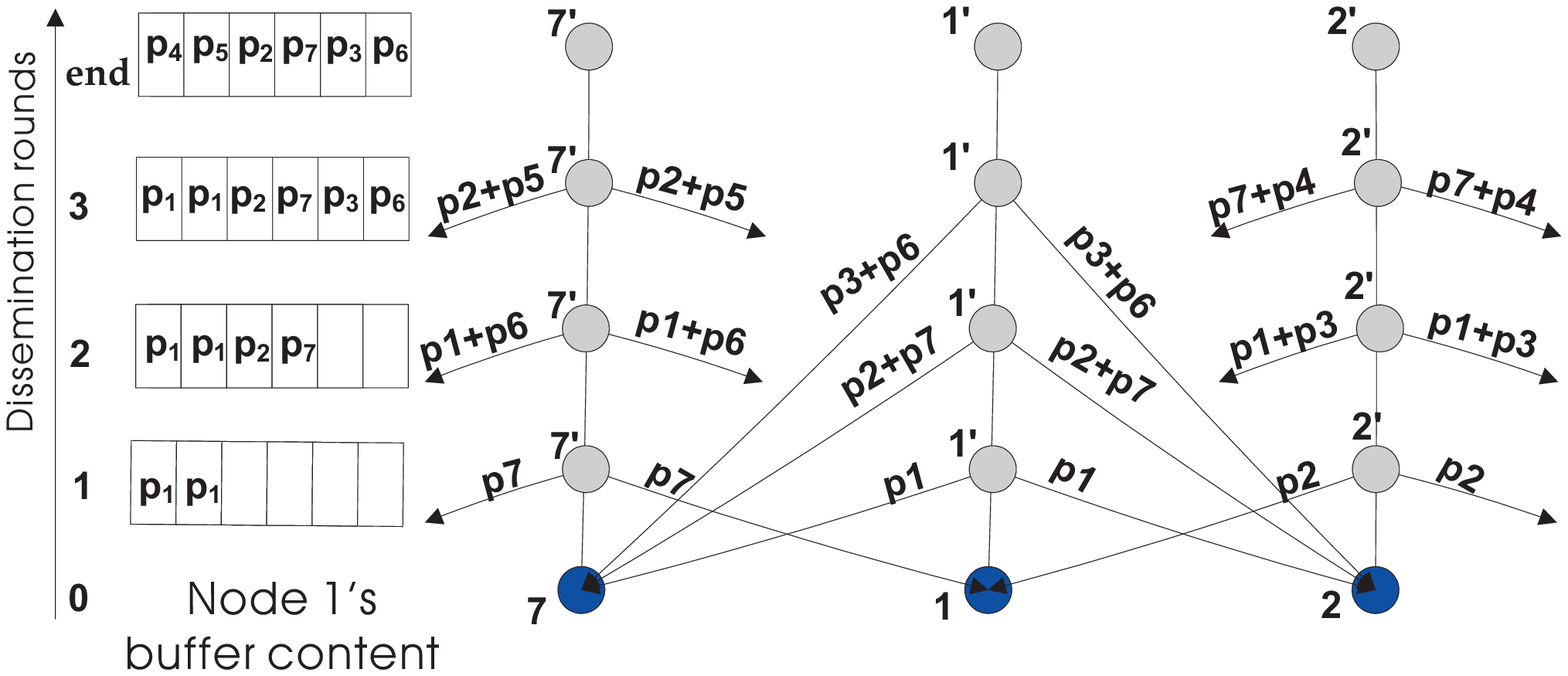}
\caption{Dissemination procedure brings all network data to each relay in half as many hops as it would be needed with simple forwarding scheme: example for $k=7$ follows the exchanges of node $1$ where the black circle on the bottom represents the node's receiver while each gray circle above it represents the transmitter at the corresponding dissemination round.}
\label{fig:dissem}
\end{center}
\end{figure}

We consider two dissemination methods: {\em no combining} in which
each relay sends its own packet and forwards each received packet
until it has seen all $k$ network packets, and {\em degree-two
combining}, described in Figure~\ref{fig:dissemalgo}.  
For the degree-two combining dissemination, a relay node combines the
packet received from its left with the packet received from its
right
 into a single packet by XOR-ing respective bits, to provide innovative information to both neighboring
 relays for the cost of one transmission~\cite{CISSBidir}. Consequently, each relay performs  a total of $\left\lceil{(k-1)/2}\right\rceil$
first-hop exchanges, as described in
Figure~\ref{fig:dissem}~and in ~\cite{WidmerFragouliLeBoudec}. Note that here storage nodes overhear degree-two packet transmissions. They either randomly combine those with previously received degree-two packets, or they first apply the on-line decoding of the packets (see Figure~\ref{fig:dissem}), and then combine obtained degree-one packets with previously stored linear combinations of degree-one packets. For further details about degree-two dissemination, the reader is referred to \cite{CissFountain}.
 \section{Decentralized Squad-Based Storage Encoding} \label{sect:Storage}
Under a centralized storage mechanism that would allow coordination
between squad nodes, a unique packet could be assigned to each of
$k$ nodes located within a supersquad of an approximate size $k/h,$ and
the same procedure repeated around the circular network for each set
of $k$ adjacent squad nodes. This periodic encoding procedure would
allow a randomly positioned IDC to collect $k$ original packets from
the set of closest nodes.
 However, our focus are scalable designs where centralized
 solutions are not possible.
 We resort to stochastic protocols for storing packet
  replicas, and apply random coding to store linear combinations of the packets.
   For each dissemination method we distinguish: {\em combining} and {\em non-combining} decentralized storage techniques.
   In both we assume that the storage squad nodes can hear (receive) any of the $k$
   dissemination transmissions from the neighboring relay nodes.
   Hence, either a common timing clock or/and regular transmission listening {\em
   is}  necessary. The reference example of non-combining (non-coding) methods is {\em coupon collection} storage, in
which each squad node randomly
 selects one of $k$ packets to store ahead of time.
As the coupon collector is completely random, it requires on average $k\log k$
storage nodes to cover all the original packets. In order to decrease the probability of many
packets not being covered, we apply combining storage techniques in
which one storage node's encoded packet contains information that
covers many original packets. The higher this {\em code symbol degree}
is, the lower is the likelihood that a packet will stay uncovered. 
We consider
 combining either {\em degree-two} or {\em degree-one} packets.  Each
squad node samples a desired code symbol degree $d$ from 
distribution $\omega(d),$ $d\in\set{1,\cdots,k}.$ The squad node
decides ahead of time which subset of $d$ transmissions it will
combine to generate the stored encoded packet. Choosing a good
distribution $\omega(d)$ is not easy, since it needs to satisfy many
contradicting requirements. The high-degree code symbols are good
for decreasing the probability of uncovered packets. However, other
requirements are more important for proper behavior of the BP decoding process, especially the right amount of degree
one and degree two code symbols. It is well known that Ideal
Soliton's {\em (IS)} expected behavior is close to ideal for Fountain codes decoded by
a BP decoder, but the large variance may cause a frequent absence
of degree-one symbols ({\em the ripple}) in the collected sample of
code symbols, thus stalling the BP process. This is the reason why
Robust Soliton {\em (RS)} is used as a choice degree distribution for rateless
erasure codes. For RS,  the probability of one-degree
symbols is overdesigned in order to prevent stalling. However,
redistribution of the probability mass from higher degrees to
degree-one increases the likelihood of uncovered packets. In the next section, we present an analysis of why IS turns out to be
better than RS when BP doping is used.
\section{Collection and Decoding} \label{sect:coldecode}
The collection problem with the coupon collector (and with similar non-combining storage methods) is straightforward as it excludes decoding. The focus is simply on providing coverage redundancy $h$ that minimizes the size of supersquad containing $k\log k$ packets required to recover $k$ source packets. For the Fountain-based combining methods, the collection problem is more elaborate, and intricately tied to decoding strategy, which we study in the following subsections. 
\subsection{Belief Propagation Decoding}
\begin{figure}[!t] 
\begin{center}
\includegraphics[width=3.7in]{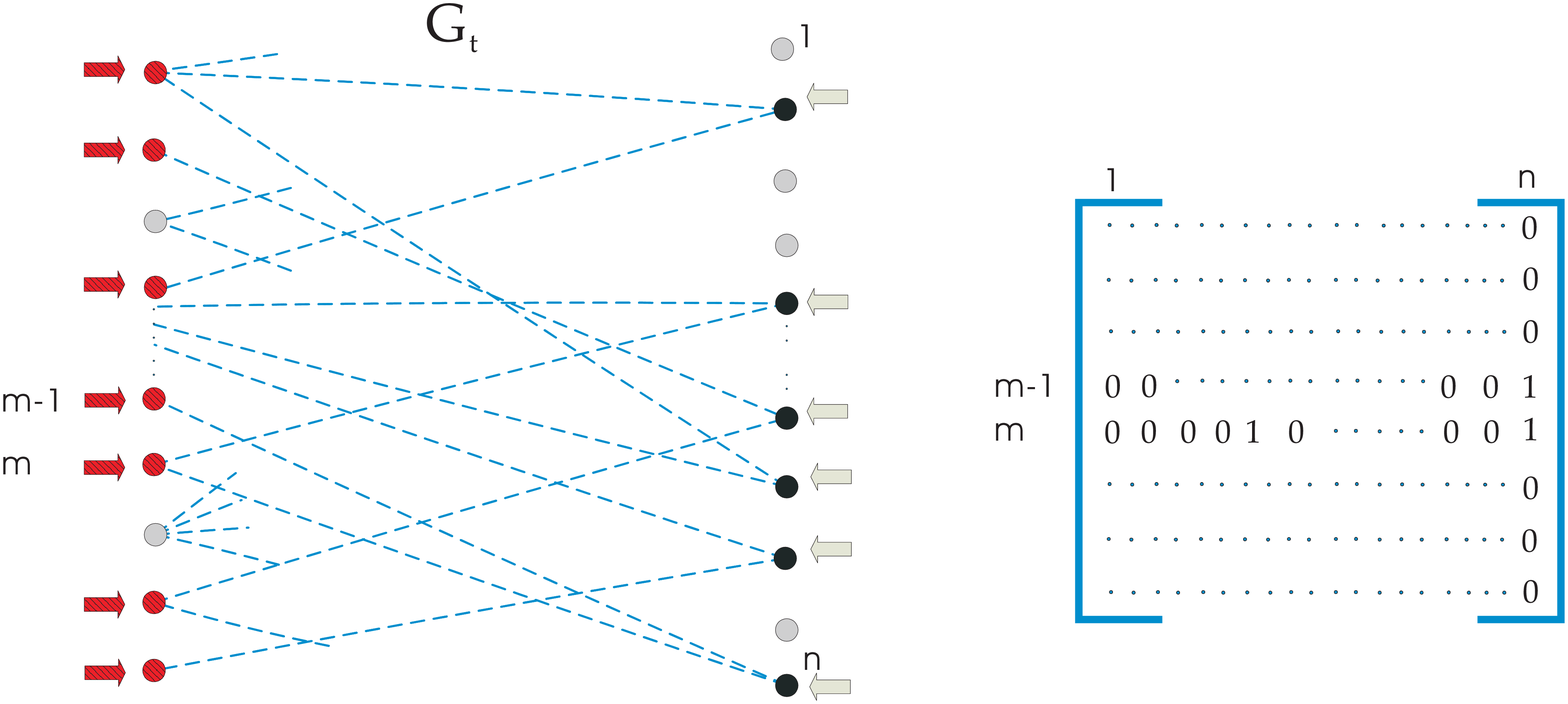}
\caption{In the graph $\mathbf{G_t},$ representing the stalled decoding process at time $t,$ we identify nodes on the left side (input symbols corresponding to rows of the incidence matrix) connected to right-hand-side nodes of degree two (output nodes corresponding to columns of weight two, represented by black nodes, and pointed to by black arrows), and then uniformly at random select one such input symbol to unlock the decoder. The set of symbols we are selecting from is represented by red nodes, indicated by red arrows.}
\label{fig:dopegraph}
\end{center}
\end{figure}

Suppose that we have a set of $k_s$ code symbols that are linear combinations of $k$ unique input symbols, indexed by the set $\curlb{1,\cdots,k}$. Let the degrees of linear combinations be random numbers that follow  distribution $\omega(d)$ with support $d\in\curlb{1,\cdots,k}$. Here, we equivalently use $\omega(d)$ and its generating
polynomial $\Omega(x)=\sum^{k}_{d=1}{\Omega_dx^d},$ where
$\Omega_d=\omega(d).$  Let us denote the graph describing the (BP)
 decoding process at time $t$ by $\mathbf{G_t}$ (see Figure~\ref{fig:dopegraph}). 
We start with a decoding matrix $\mathbf{S_0}=\set{s_{ij}}_{k\times k_s},$ where code symbols are described using columns, so that $s_{ij}=1$ iff the $j$th code symbol contains the $i$th input symbol. Number of ones in the column
corresponds to the degree of the associated code symbol. Input
symbols covered by the code symbols with degree one constitute the
ripple. In the first step of the decoding process, one input symbol
in the ripple is processed by being removed from all neighboring code symbols in the associated graph $\mathbf{G_0}$.
If the index of the input symbol is $m$, this effectively removes
the $m$th row of the matrix, thus creating the new decoding matrix
$\mathbf{S_1}=\set{s_{ij}}_{(k-1)\times k_s}.$ We refer to the code
symbols modified by the removal of the processed input symbol as
output symbols. Output symbols of degree one may cover additional input
symbols and thus modify the ripple. Hence, the distribution of
output symbol degrees changes to $\Omega_1(x)$. At each subsequent
step of the decoding process one input symbol in the ripple is
processed by being removed from all neighboring output symbols and all such output symbols that subsequently have
exactly one remaining neighbor are released to cover that 
neighbor. Consequently, the support of the output symbol degrees
after $\ell$ input symbols have been processed is
$d\in\curlb{1,\cdots,k-\ell},$ and the resulting output degree
distribution is denoted by $\Omega_{\ell}(x)$.
Our analysis of the presented BP decoding process is based on the assumption that the ripple size relative to the number of higher degree symbols is small enough throughout the process. Consequently, we can ignore the presence of defected ripple symbols (redundant degree-one symbols) \cite{HajekRipple}. Hence, the number of decoded symbols is increased by one with each processed ripple symbol.
Now, let us assume that input symbols to be processed are not taken from the ripple, but instead provided to the decoder as side information. We refer to this mechanism of processing input symbols obtained as side information as {\em doping}. In particular, to unlock the belief
propagation process stalled at time (iteration) $t,$ the degree-two 
doping strategy selects the doping symbol from the set of input
symbols connected to the degree-two output symbols in graph $\mathbf{G_t},$ as illustrated in Figure~~\ref{fig:dopegraph}. Hence, the ripple evolution is affected in a different manner, i.e. with doping-enhanced decoding process the ripple size does not necessarily decrease by one with each processed input symbol. 

The following subsections study the behavior of both varieties of the BP decoding process, first through the evolution of symbol degrees higher than one, and in particular by demonstrating the ergodicity of the Ideal Soliton degree distribution, then by modeling and analyzing the ripple process, resulting in an unified model for both classical and doping-enhanced decoding. Based on that model, we analyze the collection cost of the presented decoding strategies, when the starting $\omega(d)$ is Ideal Soliton. 
\subsection{Symbol Degree Evolution}
In this subsection, we focus on the evolution of symbol degrees higher than one (unreleased symbols), and then analyze ripple evolution separately in the next subsection. 
The analysis of the evolution of unreleased output symbols is the same for both classical BP decoding case (without doping), and the doped BP decoding. 
We now present the model of the doping (decoding) process through the column degree distribution at each decoding/doping round. We model the $\ell$th step of the decoding/doping process by selecting a row uniformly at random from the set of $(k-\ell)$ rows in the current decoding matrix $\mathbf{S_{\ell}}=\set{s_{ij}}_{(k-\ell)\times k_s}$, and removing it from the matrix. After $\ell$ rounds or, equivalently, when there are $k-\ell$ rows in the decoding matrix, the number of ones in a column is denoted by $A_{k-\ell}$. 
The probability that the column is of degree $d,$ when its length is $k-\ell-1,\ \ell\in\curlb{1,\cdots,k-3}$, is described iteratively
\begin{eqnarray}\eqnlabel{probd}
\nonumber P\paren{A_{k-\ell-1}=d}&=&P\paren{A_{k-\ell}=d}\paren{1-\frac{d}{k-\ell}}\\
&+&P\paren{A_{k-\ell}=d+1}\frac{d+1}{k-\ell}
\end{eqnarray}
for $2\leq d < k-\ell,$ and $P\paren{A_{k-\ell-1}=k-\ell}=0.$ 
Let the starting distribution of the column degrees (for the decoding matrix $\mathbf{S_0}=\set{s_{ij}}_{k\times k_s}$) be Ideal Soliton, denoted by $\rho(d),$
\begin{equation}\eqnlabel{isol}
\rho(d)=\frac{1}{d(d-1)} \mbox{\quad for $d=2,\cdots,k$},
\end{equation}
and $\rho(1)=\frac{1}{k}.$
By construction, for $l=0,$ $P\paren{A_{k}=d}=\rho(d),$ which, together with \eqnref{probd}, completely defines the dynamics of the doping process when the Fountain code is based on the Ideal Soliton.  
After rearanging and canceling appropriate terms, we obtain, for $d\geq 2,$
\begin{eqnarray}\eqnlabel{genersimple}
P\paren{A_{k-l}=d}&=&
\begin{cases}
\frac{k-l}{k}\rho(d) & d=2,\cdots,k-l,\\ 
0 & d > k-\ell.
\end{cases}
\end{eqnarray}
We assume that $k_s\approx k$ as, by design, we desire to have the set of upfront collected symbols $k_s$ as small as the set of source symbols. The probability of degree-$d$ symbols among unreleased symbols $n^{(\ell)}_u=k_s-\ell$ can be approximated with $\frac{P\paren{A_{k-\ell}=d}k_s}{k_s-\ell}\approx\frac{P\paren{A_{k-\ell}=d}k}{k-\ell}$. 
Hence, the probability distribution
$\omega_{\ell}(d)$ of the unreleased output node degrees at any
time $\ell$ remains the Ideal Soliton
\begin{equation}\eqnlabel{genersimpleU}
\omega_{\ell}(d)=\frac{k}{k-\ell}P\paren{A_{k-\ell}=d}=
\rho(d) \mbox{\quad for $d=2,\cdots,k-\ell$} .
\end{equation}

 \subsection{Doped Ripple Evolution: Random Walk Model}
\begin{figure}[!t] 
\begin{center}
\includegraphics[width=3.9in]{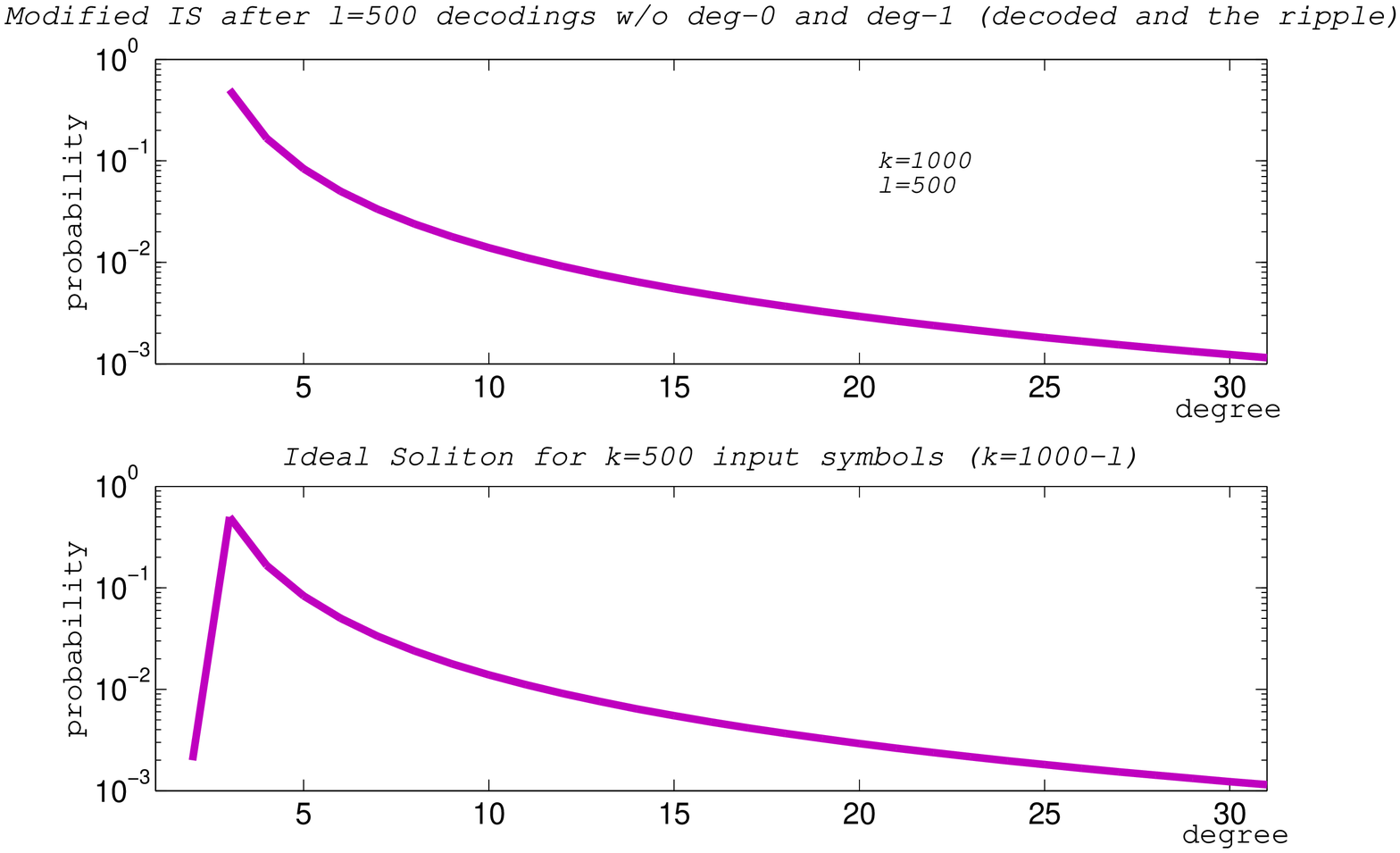}
\caption{Density Evolution of IS distribution due to uniform doping. The upper graph is the distribution of the output symbols after $\ell=500$ decodings, for initial number of collected code symbols $k=1000$; the lower graph is the IS with support set $\curlb{1,\cdots,(k=1000-\ell)}$ as if we are {\em starting} with the matrix of the same size (initial number of collected code symbols $k=1000-\ell$) as the matrix doped in round $\ell$.}
 \label{fig:dopIdensity}
\end{center}
\end{figure}

There exist comprehensive and thorough analytical models for the
ripple evolution, characterizing the decoding of LT codes
\cite{karpLTanalysis, maneva}. However, their comprehensive nature 
results in difficult to evaluate complex models. For describing the
dynamics of a doped decoder, we consider a simpler model, which
attempts to capture the ripple evolution for the Ideal Soliton.
Figure~\ref{fig:dopIdensity} and the code symbol degree evolution analysis illustrate how the Ideal Soliton distribution
maintains its shape with decoding/doping. This fact, which results
in a tractable ripple analysis and, more importantly, in an
outstanding performance as illustrated in the last section, is our
main motivator for selecting Ideal Soliton Fountain codes for our
doping scheme. 
We study the number of symbols decoded between two dopings and,
consequently, characterize the sequence of  {\em interdoping
yields.}
The time at which the $i$th doping occurs (or, equivalently, the
decoding stalls for the $i$th time) is a random variable $T_i$, and
so is the interdoping yield $Y_i=T_i-T_{i-1}.$ Our goal is to obtain
the expected number of times the doping will occur by studying the
ripple evolution. This goal is closely related to (a generalization
of) the traditional studies of the fountain code decoding which
attempt to determine the number of collected symbols $k_s$ required
for the decoding to be achieved without a single doping iteration,
i.e., when $T_1 \geq k.$

Let the number of upfront collected coded symbols be
$k_s=k\paren{1+\delta},$ where $\delta$ is a small positive value.
At time $\ell$ the total number of decoded and doped symbols is
$\ell,$ and the\ number of (unreleased) output symbols is
$n=k_s-\ell=\lambda^{\delta}_{\ell}\paren{k-\ell}.$ Here, 
$\lambda^{\delta}_{\ell}=1+\frac{k}{k-\ell}\delta$
 is an increasing
function of $\ell.$ 
The unreleased output symbol degree distribution polynomial at time
$\ell$ is $\Omega_{\ell}(x)=\sum{\Omega_{d,\ell}x^d},$ where
$d=2,\cdots,k-\ell,$ and $\Omega_{d,\ell}=\omega_{\ell}(d).$ 
In order to describe the ripple
process evolution, in the following we first characterize the
ripple increment when $\ell$ corresponds to the decoding and, next,
when it corresponds to a doping iteration.

Each decoding iteration processes a random symbol of degree-one
from the ripple.  Since the encoded symbols are constructed by
independently combining random input symbols, we can assume that the
input symbol covered by the degree-one symbol is selected uniformly
at random from the set of undecoded symbols. 
Released output symbols are its coded symbol neighbors whose output
degree is two. Releasing output symbols by processing a ripple
symbol corresponds to performing, {\em in average},
$n_2=n\Omega_{2,\ell}$ independent Bernoulli experiments with
probability of success $p_2=2/(k-\ell).$
Hence, the number of released symbols at any decoding step $\ell$ is
modeled by a discrete random variable $\Delta^{(\delta)}_{\ell}$
with Binomial distribution
$\mathbf{B}\paren{n\Omega_{2,\ell},2/(k-\ell)},$ which for large $n$
can be approximated with a (truncated) Poisson distribution of
intensity $2\Omega_{2,\ell}\lambda^{(\delta)}_{\ell}$
\begin{eqnarray}\eqnlabel{released}
\prob{\Delta^{(\delta)}_{\ell}=r}&=&(\stackrel{n_2}{r})(p_2)^r\paren{1-p_2}^{n_2-r}\\
\nonumber &\geq&\frac{(n_2)^r}{r!}(p_2)^r\paren{1-p_2}^{n_2-r}\\
\nonumber &\approx&\frac{(2\Omega_{2,\ell}\lambda^{(\delta)}_{\ell})^r}{r!}\eX{-2\Omega_{2,\ell}\lambda^{(\delta)}_{\ell}},
r=0,\cdots,n_2,
\end{eqnarray}
where we have first applied the  Stirling approximation to the
Binomial coefficient and, also, assumed that the probabilities
in ~\eqnref{released} can be neglected unless $n_2$ is much larger
than $r.$ According to \eqnref{genersimpleU}, the fraction of degree-two output
symbols for Ideal Soliton based Fountain code is expected to be
$n_2/n \approx \Omega_{2,\ell}=\rho(2)=1/2,$ for any decoding
iteration $\ell.$ Hence,
\begin{equation}\eqnlabel{releasedIdealdelta}
\prob{\Delta^{(\delta)}_{\ell}=r}=\eta(r)=\frac{\paren{\lambda^{(\delta)}_{\ell}}^r\eX{-\lambda^{(\delta)}_{\ell}}}{r!},\,\,\,  r=0,\cdots,\frac{n}{2}
\end{equation}
or, equivalently,
$\Delta^{(\delta)}_{\ell} \sim \wp\paren{\lambda^{(\delta)}_{\ell}},$ where $\wp\paren{\cdot}$ denotes Poisson distribution. 
For each decoding iteration, one symbol is taken from the ripple and
$\Delta^{(\delta)}_{\ell}$ symbols are added, so that the increments
of the ripple process can be described by random variables $X_{\ell}
= \Delta^{(\delta)}_{\ell} -1$ with the probability distribution
$\eta(r+1)$ (for $X_{\ell}=r$) characterized by the generating polynomial
$I(x)=\sum_{d=0}^{n/2}{\eta(d)x^{d-1}}$ 
and an expected
value $\lambda^{(\delta)}_{\ell}-1.$
Next we describe the ripple increment for the doping iteration, where a carefully selected input symbol is revealed at time $T_i=t_i$ when the ripple is empty (random degree-two doping).  The number of degree-two
output symbols at time $T_i=t_i$ is $n_2=\rho(2)n=n/2,$ where,
$n=\lambda^{(\delta)}_{t_i}\paren{k-t_i}.$ Degree-two doping selects
uniformly at random a row in the decoding matrix $\mathbf{S_{t_i}}$
that has one or more non-zero elements in columns of degree two.
This is equivalent to randomly selecting a
column of degree two to be released, and restarting the ripple
(i.e., same as decoding) with any of its two input symbols from the
decoding matrix whose number of degree-two columns is now $n_2-1 \approx
n_2,$ for large $n_2.$ Hence, the doping ripple increment can be
described by unit increase in addition to an
increase equivalent to the one obtained through decoding but {\em
without} the ripple decrement of $1$. That is, statistically, 
the doping ripple increment $X^D_{t_i}$ is a random variable
described by $I^D(x)=\sum{\eta(d)x^{d+1}},$ corresponding to the
shifted distribution $\eta(r-1)$ for $X^D_{t_i}=r.$

Now if, for the doping instant $t=t_{i-1},$ we define  $X_{t_{i-1}} = X^D_{t_{i-1}}-2,$ 
the ripple size for $t\in [t_{i-1}, \,\, t_i]$ can be described
in a unified manner with $S_{t,i}+2$ where
\begin{equation}\eqnlabel{ripsiz}
S_{t,i}=\sum^{t}_{j=t_{i-1}}{X_j}
\end{equation}
is a random walk modeling the ripple evolution.
 Note that the ripple increments $X_{\ell}$ are not IID random variables, since the
 intensity of $\eta(d)$ changes with each iteration $\ell$.
 However, for analytical tractability, we study the interdoping time using the random walk model
 in \eqnref{ripsiz}, by assuming that $\lambda^{(\delta)}$ changes from doping to doping,
 but remains constant within the interdoping interval. 
Under this assumption, the ripple size $S_{t,i}+2$
is a partial sum of IID random variables $X_{j},$  of the expected
value $\lambda^{(\delta)}_{t_{i-1}}-1.$ Note that, when $\delta=0,$
i.e. when $k_s=k,$ $S_{t,i}$ is a zero mean random walk. In this
special case, we treat the doping-enhanced BP process
as (an approximate) renewal process, where the process starts all
over after each doping. Modeling and analyzing this particular case
is much easier, resulting in a closed-form expression for the
expected number of dopings. We later refer to this case to provide
some intuition.
The expected interdoping yield is the expected time it takes for the
ripple random walk $S_{t,i}+ 2$ to become zero. Using random walk
terminology, we are interested in the statistics of the random-walk
stopping time. The stopping time is the time at which the decoding
process stalls, counting from the previous doping time, where the
first decoding round starts with the $0$th doping which
occurs at $T_0=0$. 
Hence, the $i$-th stopping time (doping) $T_i$ is
defined as
\begin{eqnarray} \eqnlabel{stoprule1}
T_i&=&\min\curlb{\min\curlb{t_i:S_{t,i}+2\leq 0},k}.
\end{eqnarray}
We study the Markov Chain model of the random walk  $S_{t,i}.$ Each
possible value of the random walk represents a state of the Markov
Chain {\em (MC)} described by the probability transition matrix
$\mathbf{P}_i.$ State 
$v, v\in\curlb{1,\cdots,k}$ corresponds to the ripple of size $v-1.$ State $1$
is the trapping state, with the (auto)transition probability
$\mathbf{P}_{i,11}=1$ and models the stopped random walk. Hence,
based on \eqnref{releasedIdealdelta}, we have the state transition
probabilities
\begin{eqnarray} \eqnlabel{RWtrans}
&\mathbf{P}_{i,11}&=1\\
\nonumber &\mathbf{P}_{i,v(v+b)}&=\eta(1+b),\\
\nonumber &&\mbox{ $v=2,\cdots,k$, $b=-1,\cdots,\min\paren{\left\lceil \frac{n}{2}\right\rceil,k-v}$},
\end{eqnarray}
 and $\mathbf{P}_{i,vw}=0$ otherwise, resulting in a transition probability matrix of the following almost
Toeplitz form
\begin{eqnarray} \eqnlabel{matrixRW}
\mathbf{P}_{i}&=&
\begin{bmatrix}
1 & 0 & 0 &\cdots& 0\\
\eta(0)&\eta(1)&\eta(2)&\cdots& 0\\
0&\eta(0)&\eta(1)&\cdots& 0\\
\vdots&\vdots&\vdots&\vdots&\vdots\\
0 & 0 & \cdots & \eta(0) & \eta(1)
\end{bmatrix}_{k\times k},
\end{eqnarray}
with $\eta\paren{\cdot}\equiv\wp\paren{\lambda^{(\delta)}_{t_i}}.$ 
The start of the decoding process is modeled by the MC being in the
initial state $v=3$ (equivalent to the ripple of size two). Based on
that, the probability of being in
 the trapping state, while  at time $t > T_{i},$ is
\begin{eqnarray} \eqnlabel{RWmod}
p^{(T_i)}_t=\set{0\ 0\ 1\ 0\cdots 0}\mathbf{P}_{i}^{(t-T_i)}\set{1\
0\ 0\cdots 0}^T.
\end{eqnarray}
Hence, the probability of entering the trapping state at time $t$ is
\begin{eqnarray} \eqnlabel{RWtrap}
p^{T_i}(u)&=&p^{(T_i)}_{T_i+u}-p^{(T_i)}_{T_i+u-1}\\
\nonumber &=&\set{0\ 0\ 1\ 0\cdots
0}\paren{\mathbf{P}_{i}^{u}-\mathbf{P}_{i}^{(u-1)}}\set{1\ 0\
0\cdots 0}^T,
\end{eqnarray}
where $u=t-T_i.$ 
$\curlb{T_i}$ is a sequence of
stopping-time random variables where  index  $i$ identifies a doping
round. $Y_i=T_i-T_{i-1},i>1$ is a stopping time interval of a random
walk of (truncated) Poisson IID random variables of intensity
$\lambda^{(\delta)}_{T_{i-1}}=1+\delta\frac{k}{k-T_{i-1}},$  and can
be evaluated using the following recursive probability expression
\begin{eqnarray} \eqnlabel{RecursnonIID}
\prob{Y_i=0}&=&\prob{Y_i=1}=0\\
\nonumber \prob{Y_i=t+1}&=&\eta(0)R^{\eta}\paren{t}\mbox{\quad $1\leq t<k$},\\
\nonumber R^{\eta}\paren{t}=\aleph^{(t)}(t-1)&-&\sum^{t-1}_{i=1}{\prob{Y_i=t-i}\aleph^{(i)}(1+i)}
\end{eqnarray} 
 obtained from \eqnref{RWtrap} after a series of matrix
transformations. Here, $\eta(0)$ is Poisson pdf of intensity
$\lambda^{(\delta)}_{T_{i-1}}$ evaluated at $0,$ and
$\aleph^{(s)}(d)$ is the $s$-tupple convolution of $\eta(\cdot)$
evaluated at $d,$ resulting in a Poisson pdf of intensity
$s\lambda^{(\delta)}_{T_{i-1}}$ evaluated at $d.$ The complete
derivation of \eqnref{RecursnonIID} is given in the Appendix. Note
that the intensity $s\lambda^{(\delta)}_{T_{i-1}}$ is, in general, a
random variable and that the sequence of doping times $T_i$ is a
Markov chain. Hence, the number of decoded symbols after $h$th
doping, a partial sum $D_h=\sum^h_{i=1}{Y_i}$ of interdoping yields,
is a Markov-modulated random walk.

The expected number of dopings sufficient for complete decoding is
the stopping time of the random walk $D_h,$ where the stopping
threshold is $k-u^{\delta}_k.$ Here, based on the coupon collection
model, $u^{\delta}_k$ is the expected number of uncovered symbols
(which, necessarily, have to be doped) when $k_s$ coded symbols are
collected
\begin{eqnarray} \eqnlabel{uncov}
u^{\delta}_k=k\paren{1-\frac{1}{k}}^{\set{k\paren{1+\delta}\log k}}\approx
k\mathbf{e}^{-\paren{1+\delta}\log k}.
\end{eqnarray}
The total number od dopings is the stopping time random variable $D$
defined as
\begin{eqnarray} \eqnlabel{stoprule2}
D&=&\min\curlb{h:D_{h}+u^{\delta}_k\geq k}.
\end{eqnarray}
Our model can further be simplified by replacing $T_{i-1}$ with
$l_i=\sum^{i-1}_{t=1}{\E{Y_t| T_{t-1} = l_{t}}}$ in the intensity
$\lambda^{(\delta)}_{T_{i-1}}$~\eqnref{RecursnonIID} and thus
allowing for a direct recursive computation
in~\eqnref{RecursnonIID}. 
Hence,
\begin{eqnarray}\eqnlabel{parsumnew}
\E{Y_i|T_{i-1} =
l_i}&\approx&\sum^{k-l_i}_{t=1}{t\prob{Y_i=t}}\\
\nonumber &+&\paren{1-\sum^{k-l_i}_{t=1}{\prob{Y_i=t}}}\paren{k-l_i}.
\end{eqnarray}
Furthermore, we  can approximate  $D_h\approx
l_{h+1}=\sum^h_{i=1}{\E{Y_i| T_{i-1} = l_{i}}}$ and
 use an algorithm in Figure~\ref{fig:dopingpercentagealgo} (based on \eqnref{stoprule2}) to
calculate expected number of dopings.

In special case when $\delta=0,$ further simplifying assumptions
lead to the approximation that all interdoping yields are described
by a single random variable $Y$ whose pdf is given by the following
recursive expression, based on \eqnref{RecursnonIID},
\begin{eqnarray} \eqnlabel{Recurs}
&&\prob{Y=t+1}=\\
\nonumber &&\eta(0)\paren{\wp^{(t)}(t-1)-\sum^{t-1}_{i=1}{\prob{t-i}\wp^{(i)}(1+i)}},
\end{eqnarray}
where $\wp^{(s)}(d)$ denotes Poisson distribution of intensity $s,$
evaluated at $d,$ and $t\in[0, \,\, k-1].$ The range of $t$ varies
from doping to doping, i.e. if $T_{i-1}=l_i,$ then $Y_i$ would have
support $t\in[l_i, \,\, k-1],$ and, hence, this single variable
approximation is accurate for the case when both the ripple size is
small and when $l_i \ll k.$ We now approximate the expected value of
the
interdoping yield $Y$ as 
\begin{eqnarray}\eqnlabel{parsum}
\E{Y}\approx\sum_{t=1}^k{t\prob{Y=t}} - \left(1-
\sum_{t=1}^k{\prob{Y=t}}\right) k.
\end{eqnarray}
Now, the doping process $D_h$ is a renewal process, and thus, 
the Wald Equality
\cite{Gallager-DSP} implies that the mean stopping time is $\E{D}= k/\E{Y}.$
 \section{Comparative Cost Analysis}
\begin{figure}[!t] 
\begin{center}
\includegraphics[width=3.9in]{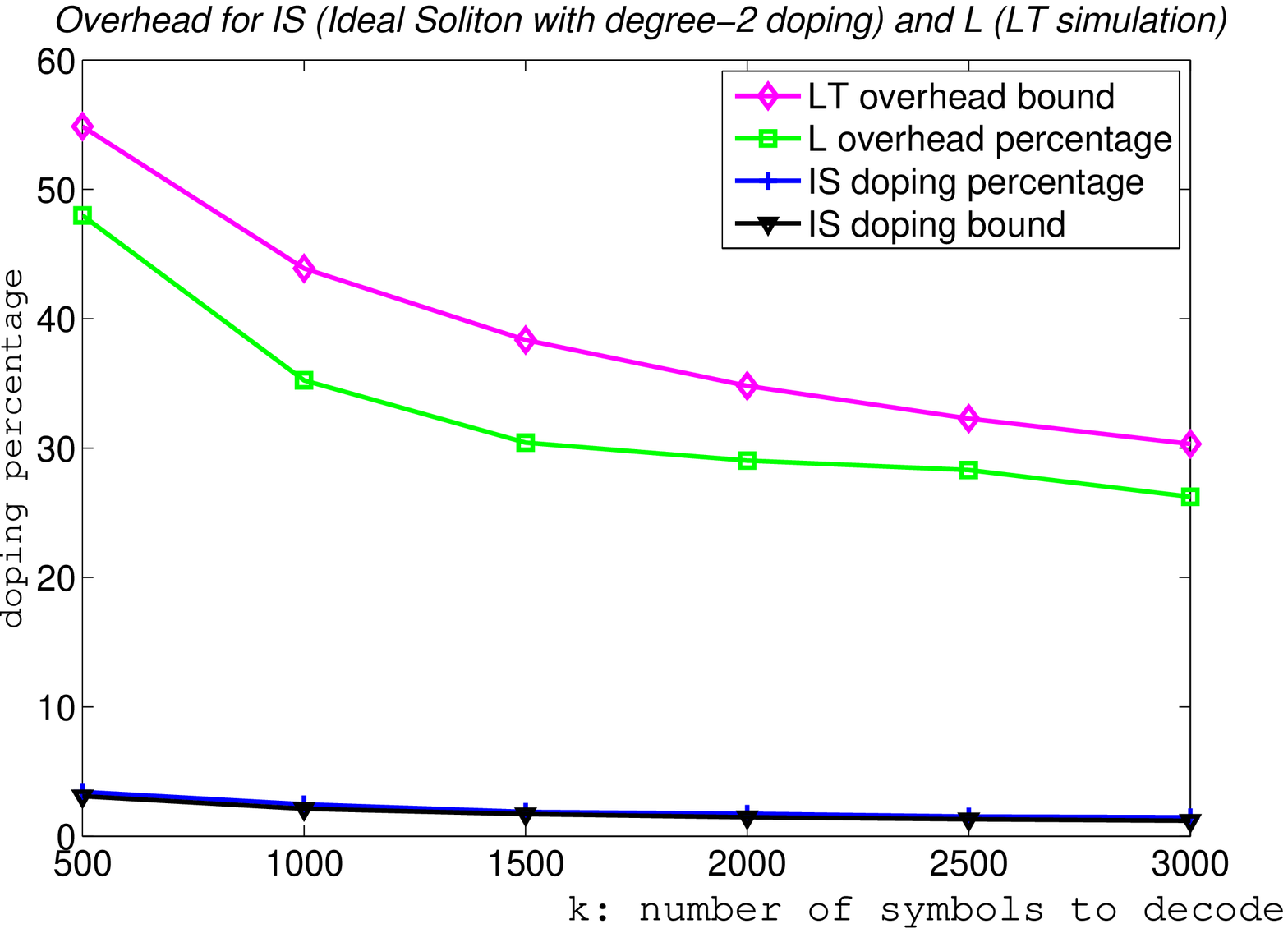}
\caption{Overhead (doping) percentage: we define $k_T(k_d)=k_s+k_d$ as the number
of symbols collected in both collection phases, and the collection
overhead ratio as $(k_T(k_d)-k)/k,$ which alows us to compare the
overhead for the simulated $LT$ decoding of $k$ original symbols and the
simulated degree-two doped belief-propagation decoding of $k$ coded symbols
with IS degree distribution. The LT overhead bound is the analytical bound by Luby~\cite{luby}.  The IS doping bound is the analytical bound based on the algorithm given in Figure~\ref{fig:dopingpercentagealgo}.}
 \label{fig:pushpull}
\end{center}
\end{figure}

\begin{figure}[!t] 
\begin{center}
\includegraphics[width=3.9in]{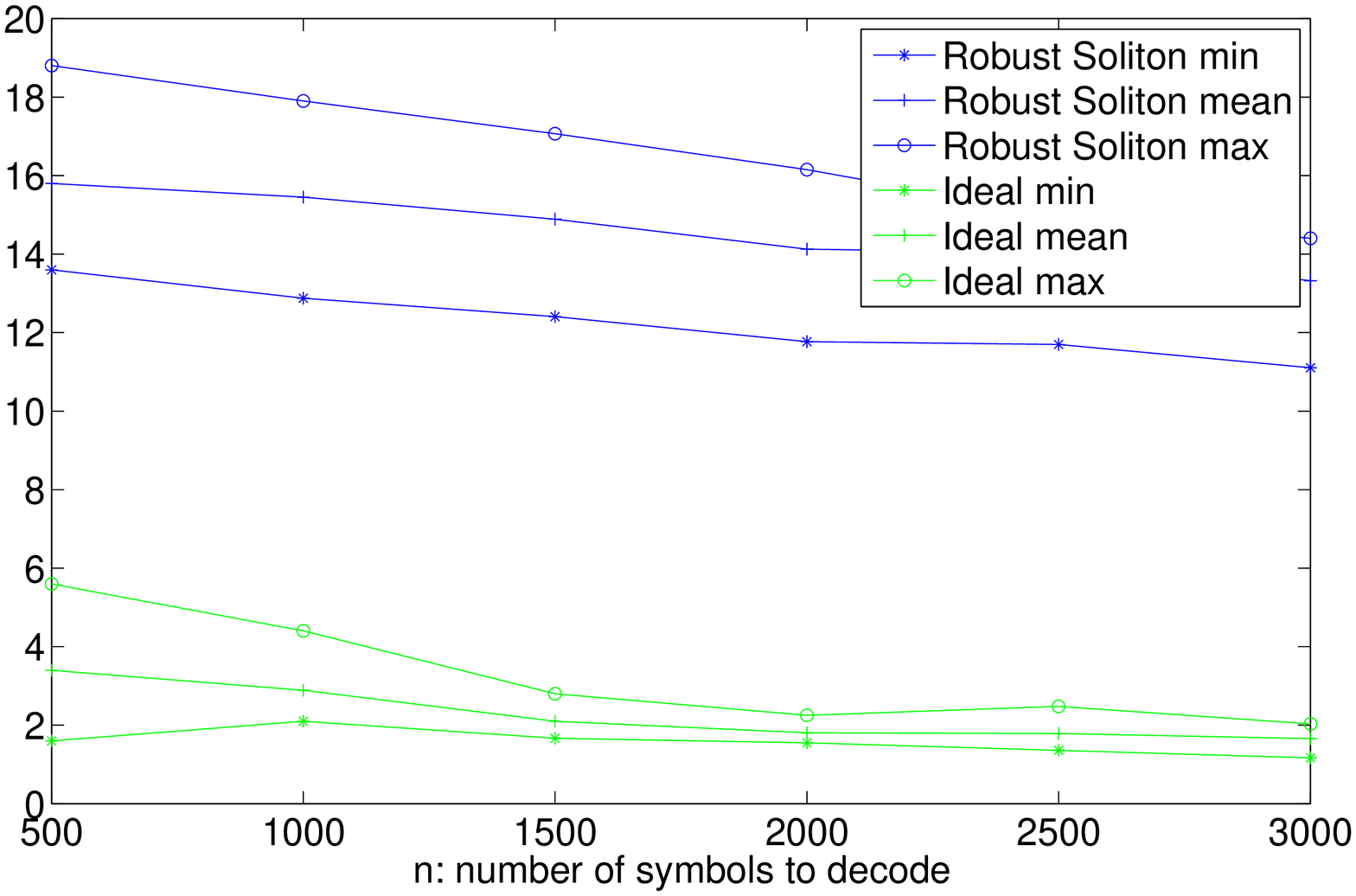}
\caption{Doping percentage
with initial IS code symbol degree distribution vs RS. Both mean and variance are much smaller for Ideal Soliton.}
\label{fig:rsideal}
\end{center}
\end{figure}

The summary of the proposed approach to
dissemination, storage, and collection with doping, based on
IS combining for storage, and a random degree-two doping
for collection strategy, is given in Figure~\ref{fig:dissStorDoping}.
We here analyze the performance of this approach in terms of data collection cost.
The cost of the upfront collection from the nearby nodes in the
super squad $1+(s-1)/4$ is significantly smaller than the collection
cost when the packets are polled from their original source relays,
which is in average $k/4.$ Nevertheless, in this section, we show
that the number of doped packets $k_d$ will be sufficiently smaller
than the residual number of undecoded symbols when the belief
propagation process first stalls, so that their collection cost is
offset, and the overall collection cost is reduced relative to the
original strategy. 
We quantify the performance of the decoding process through the
doping ratio $k_d/k.$ 
Figure~\ref{fig:pushpull} illustrates the dramatic overhead
$\paren{k_T\paren{k_d}-k}/k$ reduction when employing doping with an IS
distribution relative to the overhead of
RS encoding without doping. Figure~\ref{fig:rsideal}
demonstrates that RS with doping performs markedly worse
than IS encoding. In particular, it illustrates that
IS with doping demonstrates a very low variance, which is
surprisingly different from the results without doping.

\begin{figure}[!t] 
\begin{center}
\includegraphics[width=3.9in]{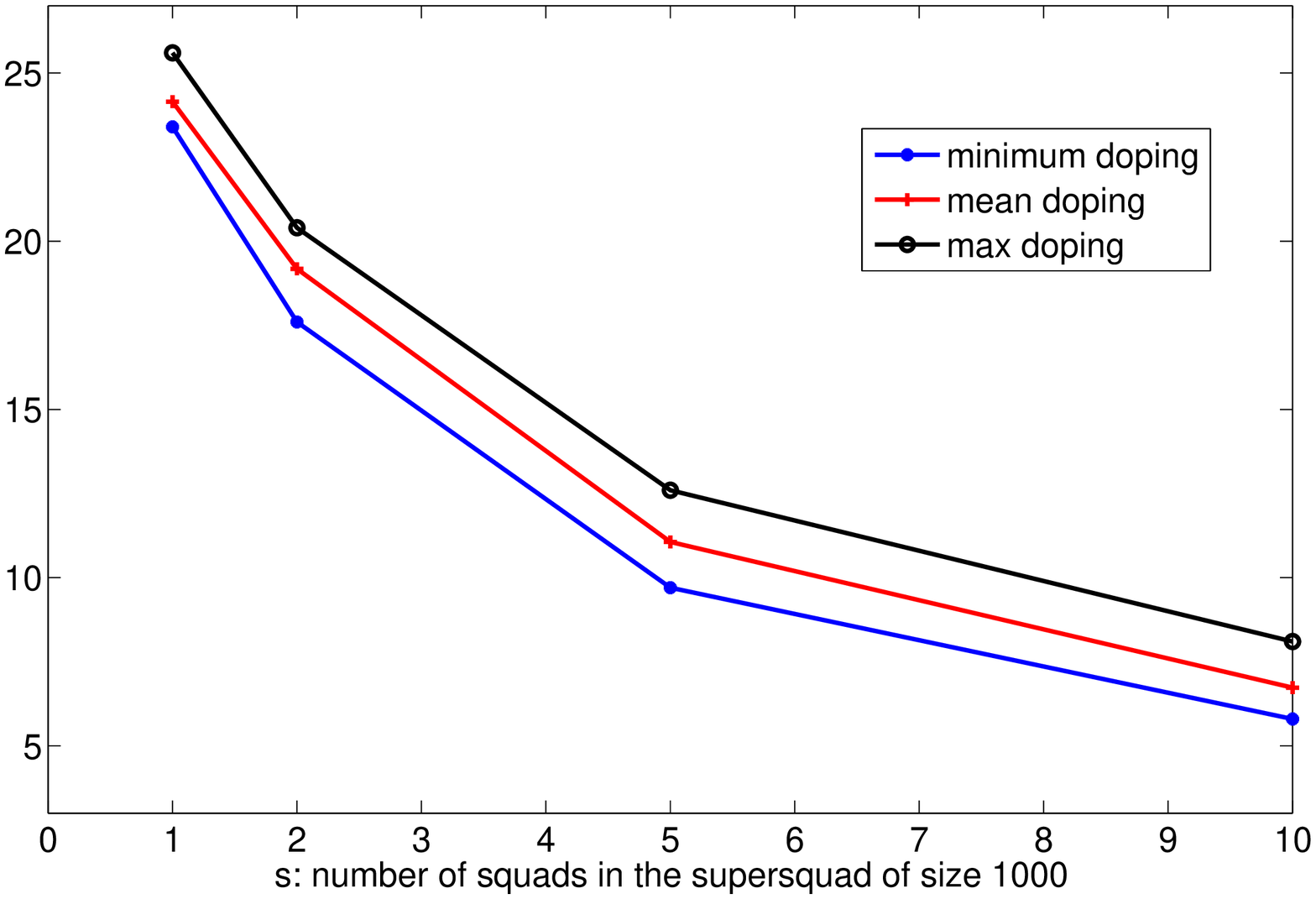}
\caption{Doping percentage as a function of supersquad size when code symbols are linear combinations of degree-two packets: for a fixed number of upfront collected symbols $k_s=1000$, encoded by
degree-two IS method, the squad size (node density) is
changed, so that the supersquad contains $1,2,5,$ and $10$ squads.
The more squads there are, the more intense is the data mixing,
decreasing the probability of non-covered original symbols.}
\label{fig:deg2b}
\end{center}
\end{figure}

\begin{figure}[!t] 
\begin{center}
\includegraphics[width=3.9in]{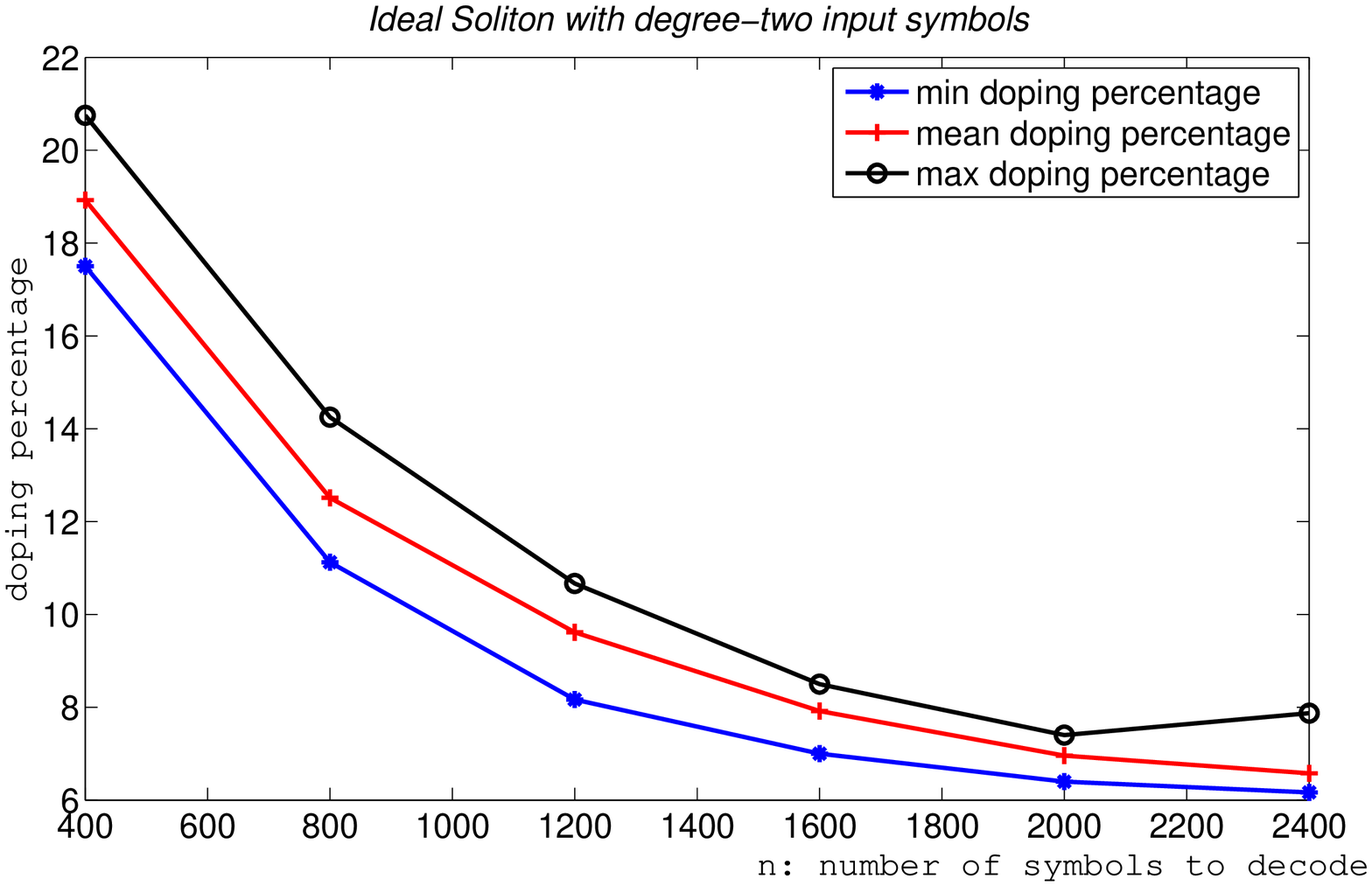}
\caption{The encoding process emulates supersquads with
fixed squad size $h=200$ and the degree-two input symbols overheard
within the superquad: the resulting doping percentage for IS degree distribution of stored code symbols.}
\label{fig:deg2a}
\end{center}
\end{figure}

In Section~\ref{Sect:Motiv} we characterized a circular squad network by its node density $\mu$ and its source density $\mu_s$ so that network scaling can be expressed through the scaling of the coverage redundancy factor $h=\mu/\mu_s.$
Figure~\ref{fig:deg2b} illustrates the importance of considering coverage redundancy when selecting storage/ collection strategy: in the case of {\em degree-two}
dissemination,  when the size of the supersquad $s$ increases the fountain code strategy improves (in terms of a
reduced doping $k_d/k$ required for decoding) due to an increase in
mixing. As a result of the mechanism described in Figures~\ref{fig:dissem}~and~\ref{fig:dissemalgo}, the degree-two combinations in adjacent squads have a significant number of common packets. Hence, when forming code symbols by combining degree-two packets, we encounter a higher code symbol dependency and an increased number of redundant symbols in the ripple, which increases the probability of uncovered input symbols. Our doping overhead accounts for uncovered symbols, since ultimately they need to be pulled off the original sources for the complete data recovery. With increased supersquad size $s$ (or, equivalently, a decreased $h$) the mixing of input symbols is improved and this negative effect is alleviated.  
This
dependency is not present in the case of degree-one dissemination. 
Figure~\ref{fig:deg2a} gives the corresponding required doping
$k_d/k$ as a function $k$ for a fixed squad size $h=200.$

\begin{figure}[!t] 
\begin{center}
\includegraphics[width=3.9in]{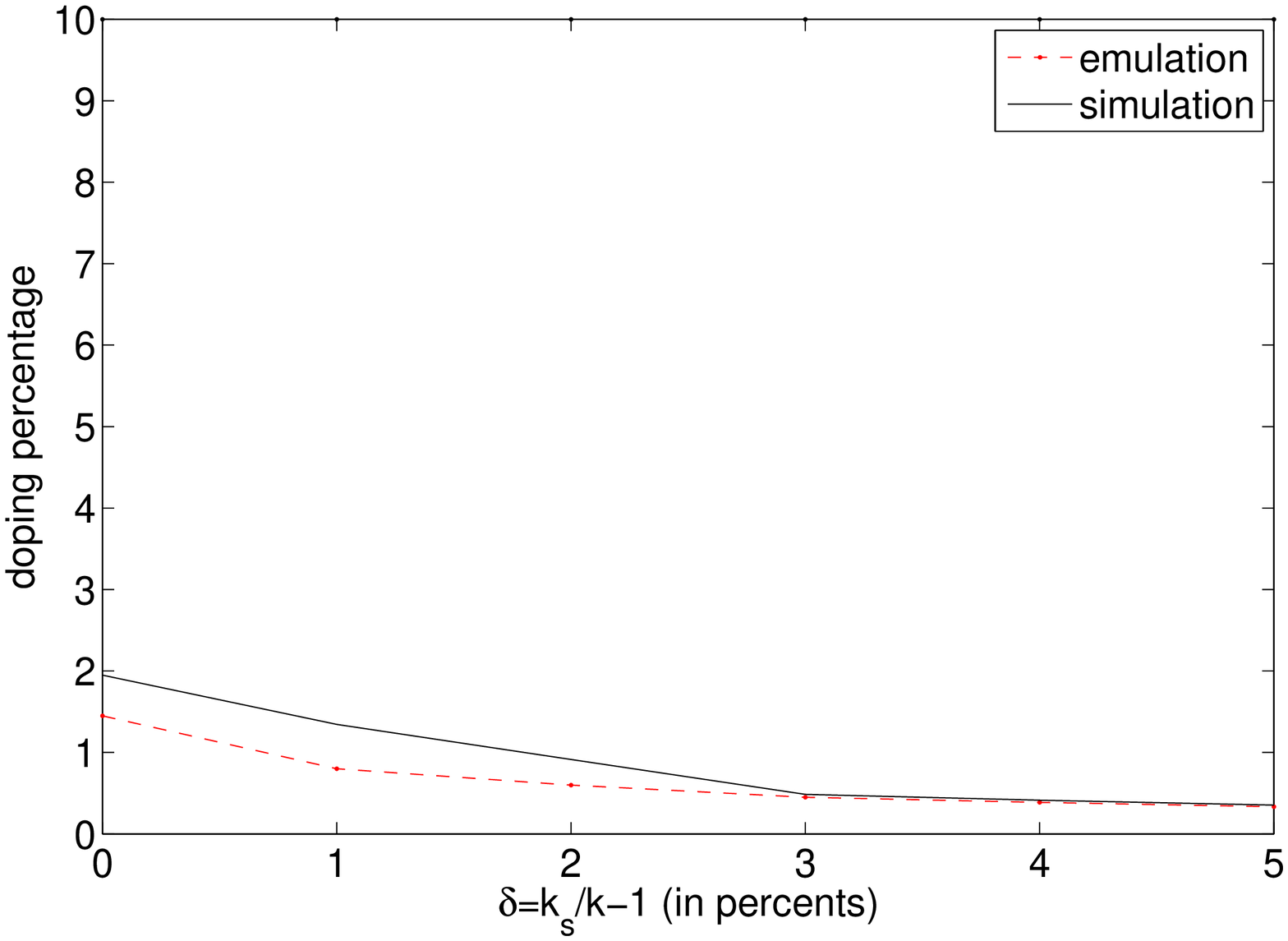}
\caption{Doping percentage for different values of $\delta=k_s/k-1.$ Emulation results are obtained based on our analytical model and algorithm in Figure~\ref{fig:dopingpercentagealgo} }
\label{fig:doperange}
\end{center}
\end{figure}

\begin{figure}[!t] 
\begin{center}
\includegraphics[width=3.9in]{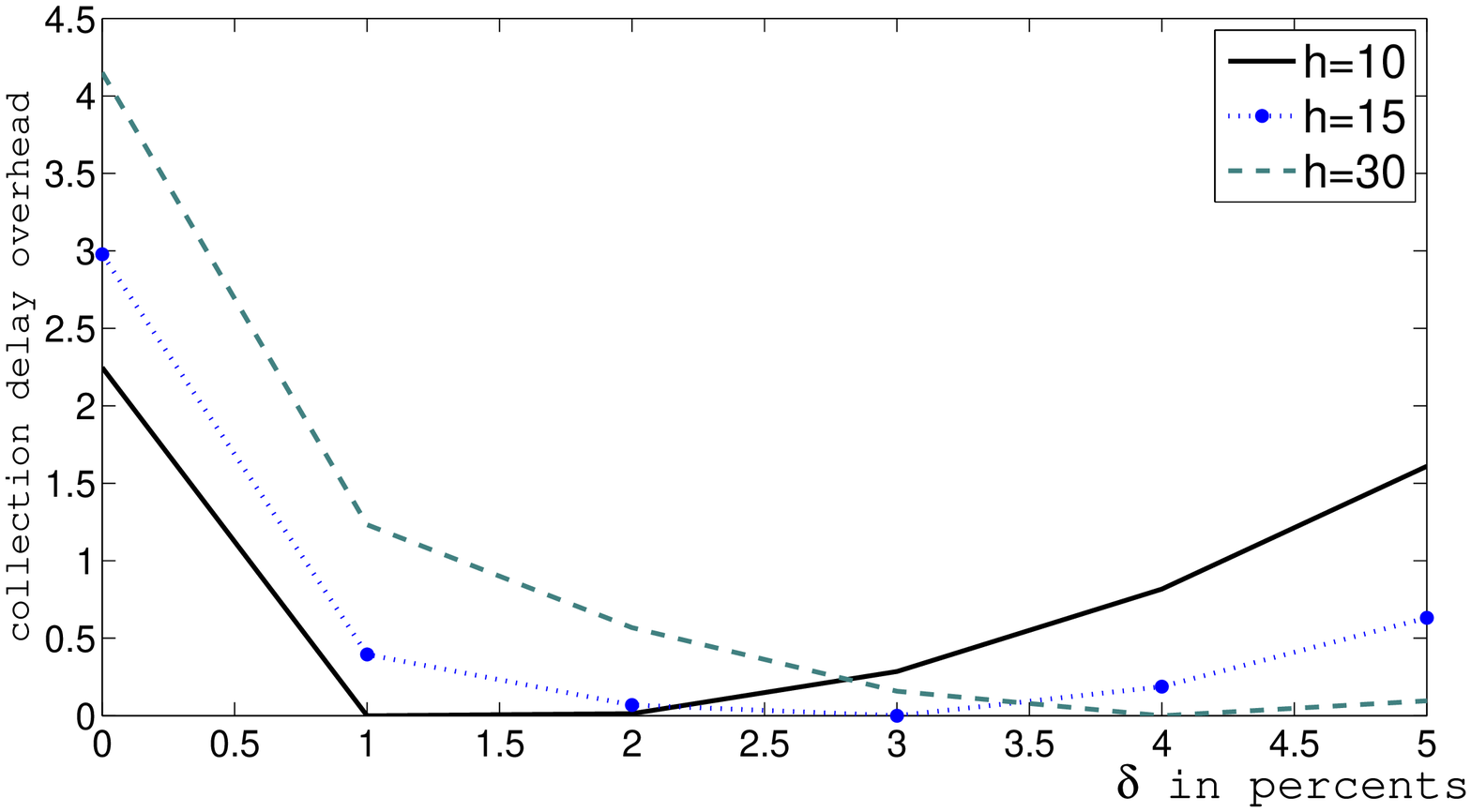}
\caption{Collection delay (hop count) above minimum per input symbol for different values of coverage redundancy $h$ as a function of $\delta$. Note that there is an optimal $\delta$ for each $h$ in which the delay is minimized: for $h=10$ $\delta$ is one percent, for $h=15$ it is $3\%$ percent, for $h=30$ $\delta=4\%$}
\label{fig:optdopa}
\end{center}
\end{figure}

\begin{figure}[!t] 
\begin{center}
\includegraphics[width=3.9in]{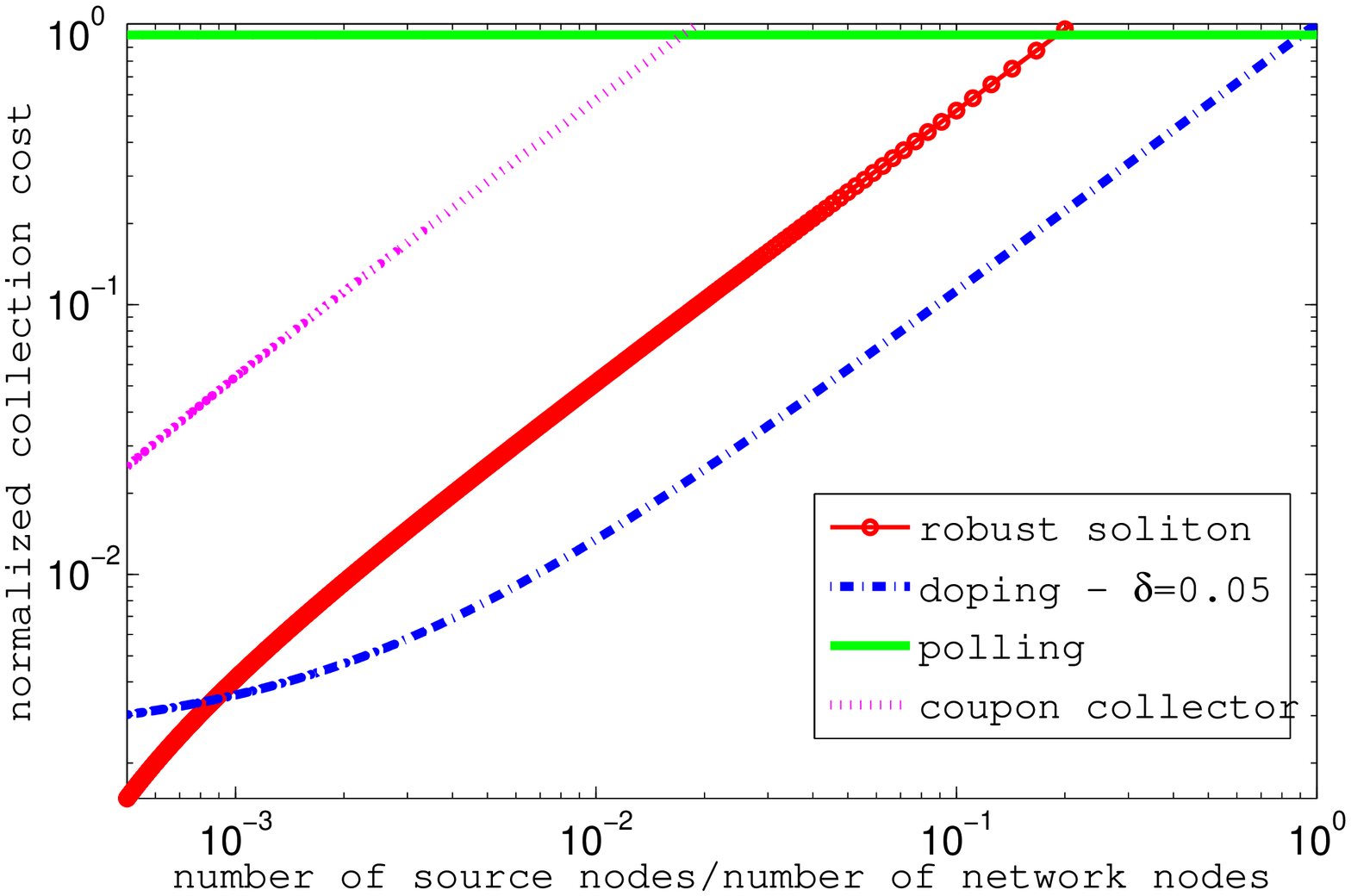}
\caption{Collection Delay for various collection techniques,
normalized with respect to the polling cost, as a function of $1/h$. Note that the proposed doping strategy is inferior to polling only when there are no other nodes but relays. For very large squads ($>1000$), the proposed doped IS code induces a sufficiently large polling cost (usually to start the process as the IS sample is likely not to have degree-one symbols) which
offsets (and exceeds) the cost due to overhead packets solicited from the
supersquad with the RS-based strategy without doping. The
coupon collection (non-coding) strategy is consistently worse by an order
of magnitude than the RS-based fountain encoding and is worse
than polling for high source densities (small squads with tens of nodes).} 
\label{fig:optdopb}
\end{center}
\end{figure}

%
The {\em cost minimization problem} for any encoding scheme with (and
without) doping is described as follows. Let, the pair $(k_s,k_d)$
be the feasible number of encoded and doped packets when sufficient
for decoding the original $k$ packets. The per-source packet
collection cost for this pair is
\begin{eqnarray}\eqnlabel{costeq}
c_T(h) = \set{c_s(h)k_s + c_d k_d}/k
\end{eqnarray}
where $c_s(h) = 1+(s(h)+1)/4$ is the average collection cost from the
supersquad of size $s(h) = \left\lceil k_s/h\right\rceil$ and
$c_d=\left\lceil k/4\right\rceil$ is the average collection doping cost when
polling doped packets from the original source relays. Examples of
$(k_s,k_d)$ pairs are $(0,k)$ for the pure polling mechanism with
cost $c_T(h)=c_d=\left\lceil k/4\right\rceil $ and $(k_s=
k+\sqrt(k)\log^2(k/\delta),0)$ in average for  degree-one
dissemination and RS fountain encoding with average
per-packet cost $c_T=c_s(h)k_s/k.$ For any given encoding mechanism
and the set of feasible pairs $(k_s,k_d)$, the minimum per-packet collection cost is
$c_{min}(h)=\min_{(k_s,k_d)} c_T(h).$
The effect on the doping percentage of increasing the number of
upfront collected symbols $k_s$ above $k$ (described by our general model
of interdoping times) is illustrated in Figure~\ref{fig:doperange}. 
Figure~\ref{fig:optdopa} illustrates per-packet collection cost above minimum, based on ~\eqnref{costeq}, as a function of the number of 
packets $(k_s-k)/k$ collected from the supersquad in excess of $k,$ for different values of coverage redundancy $h,$  and IS encoding. For the range of coverage redundancies that may be of practical value (up to 50), the minimum collection cost is obtained for
$k_{s,min}/k\in(1,1.05).$
Figure~\ref{fig:optdopb} illustrates the per-packet cost
$c_{T}(h)/(k/4)$ normalized to the reference polling cost as a
function of $\lambda_s/\lambda = 1/h,$ the relative density of source
nodes for a network with $k=2000$ source packets. Four strategies
are included all based on degree-one packet dissemination: reference
polling, degree-one coupon collection, RS with no doping,
and the IS encoding with a feasible doping pair $(k_s,k_d).$ Note that the proposed scheme is inferior to the RS-based scheme only for very low density of events, i.e. when $h>1000.$

In conclusion, in this paper we showed that, for the circular squad network, the total
collection cost could be reduced by applying a packet combining
degree distribution that is congruous to doping, applying a good doping mechanism, and by balancing
the cost of upfront collection and doping, given coverage redundancy factor. The proposed network model that includes a route of relays and the nodes overhearing relays' transmissions is chosen based on a range of sensor network applications that monitor physical phenomena with linear spatial blueprint, such as road networks and border-security sensor nets. In order to limit the scope of the paper, we here omit describing a more general setup in which our network model can be used. However, we argue that networks of different (non-linear) topology may use dissemination mechanisms that produce shortest routing paths from data sources to the collection node, suggesting a cost collection analysis based on these "linear route networks" and, hence,  similar to the one presented here. This is one of the reasons we treat data dissemination separately from data collecting in our cost analysis.

\bibliographystyle{plain}
\bibliography{journalNC2col}
\section*{Appendix}
{\subsection* {Random Walk Ripple Evolution: The Stopping Time
Probability}} Recall that for the Markov Chain model of the ripple
evolution, described by \eqnref{RWtrans} and \eqnref{matrixRW}, its
trapping state corresponds to the empty ripple. The probability of
entering the trapping state at time $t,$ where $t>T_{i},$  is given in
\eqnref{RWtrap}, where $u=t-T_i.$
The probability of being in the trapping state at $T_i+u$ can also
be expressed as $p^{(T_i)}_{T_i+u}=\set{0\ 0\ 1\ 0\cdots
0}\mathbf{P}_{i}^{(u-1)}\set{1\ \eta(0)\ 0\cdots 0}^T.$ Hence, we
can reformulate \eqnref{RWtrap} as
\begin{eqnarray} \eqnlabel{RWtrappe}
p^{T_i}(u)&=&\set{0\ 0\ 1\ 0\cdots 0}\mathbf{P}_{i}^{u-1}\set{0\ \eta(0)\ 0\cdots 0}^T.
\end{eqnarray}
Note that both $\set{0\ 0\ 1\ 0\cdots 0}$ and $\set{0\ \eta(0)\
0\cdots 0}^T$ have zero-valued  first elements, which means that the
first row and the first column of the transition probability matrix
$\mathbf{P}_{i}$ do not contribute to the value of
\eqnref{RWtrappe}. Hence, we introduce a new matrix
$\mathbf{\widetilde{P}}_{i}$ which contains the significant elements
of $\mathbf{P}_{i}$ as
\begin{eqnarray} \eqnlabel{matrixRWsignif}
\mathbf{\widetilde{P}}_{\ell}&=&
\begin{bmatrix}
\eta(1)&\eta(2)&\eta(3)&\cdots& 0\\
\eta(0)&\eta(1)&\eta(2)&\cdots& 0\\
0&\eta(0)&\eta(1)&\cdots& 0\\
\vdots&\vdots&\vdots&\vdots&\vdots\\
0 & 0 & \cdots & \eta(0) & \eta(1)
\end{bmatrix}_{k-1\times k-1},
\end{eqnarray}
whith $\eta\paren{\cdot}\equiv\wp\paren{\lambda^{(\delta)}_{\ell}}.$
Now,
\begin{eqnarray} \eqnlabel{RWalgo}
p^{T_i}(u)&=&\eta(0)\set{0\ 1\ 0\cdots
0}\mathbf{\widetilde{P}}_{i}^{(u-1)}\set{1\ 0\ 0\cdots 0}^T.
\end{eqnarray}
Assuming $n$ is large, we can approximately express the $u$th power
of the matrix $\mathbf{\widetilde{P}}_{i}$ through a matrix that
contains elements ${\aleph^{(u)}( ) }$ of the $u$th convolution of
the pdf array $\overline{\eta}=\set{\eta(0)\ \eta(1)\ \cdots}.$ Let
us define $\overline{\eta}$ as degree-one convolution. For order-two
convolution, we convolve $\overline{\eta}$ with itself, and
 $u$th convolution of $\overline{\eta}$ is obtained by
recursively convolving $(u-1)$th convolution with $\overline{\eta}.$
By multiplying the matrix
\begin{eqnarray} \eqnlabel{firstconvmat}
\mathbf{\widetilde{P}}^C_{i}&=&\begin{bmatrix}
\eta(0)&\eta(1)&\eta(2)&\cdots\\
0&\eta(0)&\eta(1)&\cdots\\
\vdots&\vdots&\vdots&\vdots&\\
0 & \cdots & \eta(0) & \eta(1)
\end{bmatrix},
\end{eqnarray} which was obtained by adding the column $\set{\eta(0)\ 0\ 0\ \cdots}^T$ in front of $\mathbf{\widetilde{P}}_{i}$, and another matrix
\begin{eqnarray} \eqnlabel{secondconvmat}
\mathbf{\widetilde{P}}^R_{i}&=&\begin{bmatrix}
\eta(2)&\eta(3)&\eta(4)&\cdots\\
\eta(1)&\eta(2)&\eta(3)&\cdots\\
\eta(0)&\eta(1)&\eta(2)&\cdots\\
0&\eta(0)&\cdots&\cdots\\
\vdots&\vdots&\vdots&\vdots
\end{bmatrix},
\end{eqnarray} which was obtained by adding the row $\set{\eta(2)\ \eta(3)\ \eta(4)\cdots}$ above $\mathbf{\widetilde{P}}_{i},$ we obtain
\begin{eqnarray} \eqnlabel{convmat}
\mathbf{\widetilde{P}}^C_{i}\mathbf{\widetilde{P}}^R_{i}&=&\begin{bmatrix}
\aleph^{(2)}(2)&\aleph^{(2)}(3)&\cdots\\
\aleph^{(2)}(1)&\aleph^{(2)}(2)&\cdots\\
\aleph^{(2)}(0)&\aleph^{(2)}(2)&\cdots\\
\cdots&\cdots&\cdots&\cdots
\end{bmatrix}\\
&=&\mathbf{\widetilde{D}}^{(2)},
\end{eqnarray}
where $\aleph^{(s)}(d)$ is the $s$-th convolution of $\eta(\cdot)$
evaluated at $d,$ and $\mathbf{\widetilde{D}}^{(2)}$ is what we
refer to as second convolution matrix of $\overline{\eta},$ for
$\eta\paren{\cdot}\equiv\wp\paren{\lambda^{(\delta)}_{t_i}}.$ Hence,
\begin{eqnarray} \eqnlabel{powermat}
{\mathbf{\widetilde{P}}_{i}}^2&=&
\mathbf{\widetilde{D}}^{(2)}-\eta(0)
\begin{bmatrix}
\eta(2)&\eta(3)&\cdots\\
0&0&\cdots\\
\vdots&\vdots&\vdots
\end{bmatrix}\\
\nonumber &=&\mathbf{\widetilde{D}}^{(2)}-\set{\eta(0)\ 0\
0\cdots}^T\set{\aleph^{(1)}(2)\ \aleph^{(1)}(3)\
\aleph^{(1)}(4)\cdots}.
\end{eqnarray}

\noindent By induction,
\begin{eqnarray}
{\mathbf{\widetilde{P}}_{i}}^3 &=& \mathbf{\widetilde{D}}^{(3)} -\set{\eta(0)\ 0\ 0\cdots}^T\set{\aleph^{(2)}(3)\ \aleph^{(2)}(4)\cdots}\\
\nonumber &-&\mathbf{\widetilde{P}}_{i}\set{\eta(0)\ 0\cdots}^T\set{\aleph^{(1)}(2)\ \aleph^{(1)}(3)\cdots},\\
\nonumber {\mathbf{\widetilde{P}}_{i}}^u &=&
\mathbf{\widetilde{D}}^{(u)}-\sum^{u}_{z=2}{{\mathbf{\widetilde{S}}_{i}}^u(z) }\\
\nonumber {\mathbf{\widetilde{S}}_{i}}^u(z)&=&{\mathbf{\widetilde{P}}_{i}}^{(u-z)}\set{\eta(0)\
0\cdots}^T\set{\aleph^{(z-1)}(z)\ \aleph^{(z-1)}(z+1)\cdots}.\eqnlabel{powerind}
\end{eqnarray}
Replacing \eqnref{powermat} in  \eqnref{RWalgo}, we obtain  \eqnref{RecursnonIID}.


\begin{figure}[t]
\begin{flushleft}
\begin{description}\small
\item [\bf Initialization:]
\item[\ \ k=1:]{\small 
Relay $i$ sends its own packet $p_i$, and subsequently receives the
packets $p_{(i-1)}$ and $p_{(i+1)}$ originating from its first-hop
neighbors.}
\item[\ \ k=2:]{\small 
Relay $i$ sends a linear combination (XOR) of the received packets
$p_{(i-1)}$ and $p_{(i+1)}$, and subsequently receives the packets
containing $p_i$ XOR-ed with the packets $p_{(i-2)}$ and $p_{(i+2)}$
originating from its second-hop neighbors, respectively. Relay $i$
recovers $p_{(i-2)}$ and $p_{(i+2)}$ by XOR-in the received linear
combinations with $p_i$.}

\item [\bf For $\paren{k=3, k<(n+1)/2, k++}$]

    \item[\ \ {\bf Online Decoding}] 
    \item[] The packets received by relay $i$ in the $(k-1)$th round contain
    linear combination of packets $p_{(i-k+2)}$ and $p_{(i+k-2)}$ and
    packets $p_{(i-k+1)}$ and $p_{(i+k-1)}$, originating from its $(k-1)$th hop neighbors.
    XOR-ing the received packets with the matching packets $p_{(i-k+2)}$ and $p_{(i+k-2)}$,
    the relays recover the packets $p_{(i-k+1)}$ and $p_{(i+k-1)}$.
    \item[\ \ {\bf Storing}] 
    \item[] The buffer space is updated with the recovered original packets $p_{(i-k+1)}$ and $p_{(i+k-1)}$.
    For $k>3$ the buffer space is updated by overwriting packets $p_{(i-k+4)}$ and $p_{(i+k-4)}$.
    \item[\ \ {\bf Encoding}] 
    \item[] In the $k$th round, relay $i$ linearly combines packets $p_{(i-k+1)}$ and $p_{(i+k-1)}$,
    and transmits the linear combination.
\end{description}
\end{flushleft}
\caption{Degree-two Dissemination Algorithm}
\label{fig:dissemalgo}
\end{figure}

\begin{figure}[t]
\begin{flushleft}
\begin{description}\small
\item [\bf Dissemination and Storage:]
\item[\ \ ]{\small degree-one/two  dissemination  of $k$  source packets; each storage node stores a random linear combination of $d$ disseminated packets; $d$ is drawn from
 IS $\rho(d)$.}
\item [\bf Upfront collection:]
\item[\ \ ]{\small IDC collects $k_s$ encoded packets from $s$ closest storage squads.}
\item [\bf Belief propagation decoding and doping-collection:]
\item[\ \ ]{\small $l=0$: number of processed source packets }
\item[\ \ ]{\small $k_{r,l}$: number of packets in the ripple }
\item[\ \ ]{\small $k_d=0$: number of doped packets.}
\item [\ \bf For $\paren{l=0, l\leq k, l++}$]
\item[\ \ ]{{\bf while} $k_{r,l} = 0$}
\item[\ \ \ ] Collect(from the source relay) and dope the decoder with a source packet contributing to
a  randomly selected degree-two (or larger) output packet.
\item[\ \ \ ] $k_d++; l++;$
\item[\ \ ]{\bf endwhile}%
\item[\ ]{\small Process a symbol from the ripple; $k_{r,l}--;$}
\item[\ \bf endfor]
\end{description}
\end{flushleft}
\caption{Proposed dissemination, storage, and doping collection}
\label{fig:dissStorDoping}
\end{figure}

\begin{figure}[t]
\begin{flushleft}
\begin{description}\small
\item [\bf Initialization:]
\item[\ \ ]$l_i=0, D=0$

\item [\bf For $\paren{i=1, D<k, i++}$]

\item[\ \ ]Calculate $\lambda^{(\delta)}\paren{l_i}$
\item[\ \ ]Using \eqnref{RecursnonIID}, calculate $\prob{Y_i=t}$ for $t\leq k-l_i$
\item[\ \ ]Using \eqnref{parsumnew}, calculate $\E{Y_i}$
\item[\ \ ]$D=D+\E{Y_i}$
\item[\ \ ]$l_i=D$
\item [\bf $k_d=i, p_d=100k_d/k$]
\end{description}
\end{flushleft}
\caption{Calculation of the expected doping percentage $p_d$ based on the number of upfront collected symbols}
\label{fig:dopingpercentagealgo}
\end{figure}
\end{document}